\documentclass[prb,twocolumn,superscriptaddress,preprintnumbers,amsmath,amssymb]{revtex4-2}
\usepackage{graphicx}
\usepackage{dcolumn}
\usepackage{bm}
\usepackage{color}
\usepackage{physics}
\usepackage{tikz,xcolor,hyperref}
\usepackage{lipsum}
\usepackage[T1]{fontenc}
\usepackage{booktabs}
\usepackage{multirow, array}
\usepackage{vwcol}
\usepackage{float}
\usepackage{ulem}
\usepackage{threeparttable}
\usepackage{siunitx}
\usepackage{nicematrix}
\usepackage{makecell}
\usepackage[version=4]{mhchem}
\usepackage{cleveref}
\usepackage{gensymb}
\usepackage[caption=false]{subfig} %
\newcommand{\be}{\begin{equation}}
\newcommand{\ee}{\end{equation}}
\newcommand{\bea}{\begin{eqnarray}}
\newcommand{\eea}{\end{eqnarray}}

\hypersetup{colorlinks=true,linkcolor=blue,citecolor=blue}

\newcommand{\orcidlink}[1]{\href{https://orcid.org/#1}{\includegraphics[width=10pt]{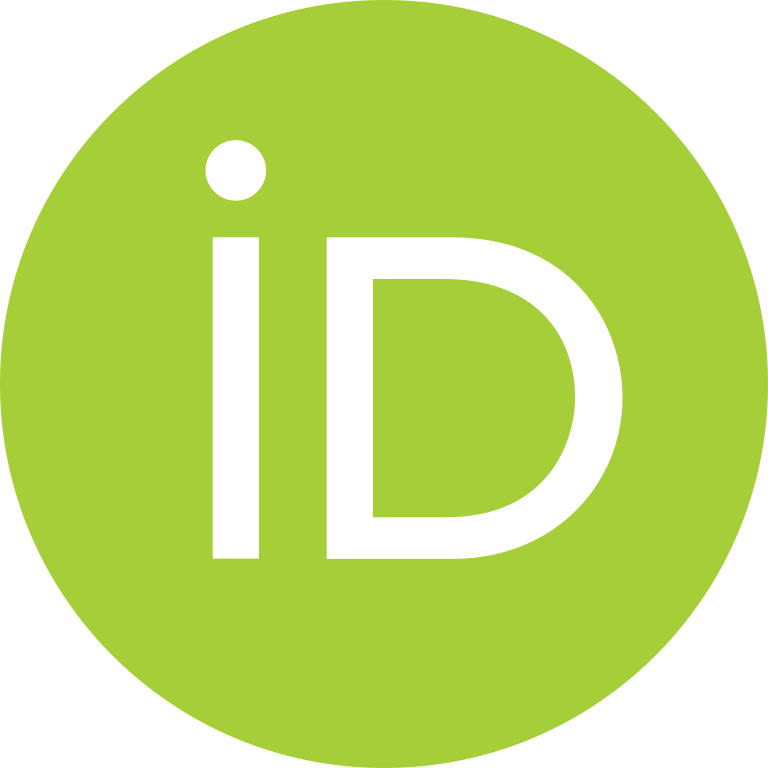}}}

\begin{document}

\title{Rashba Effect and Type--III Broken--Gap Band Alignment in MoTe$_2$/HfS$_2$ van der Waals Heterostructures for Tunnel Field--Effect Transistors}

\author{Daniel S. Lopes\orcidlink{0000-0002-5075-1334}}
\email{daniel.lopes@fisica.ufc.br}
\affiliation{Department of Physics, University of Cear\'a, Fortaleza, CE, Brazil}
\affiliation{Institute of Physics, University of Mato Grosso, Cuiab\'a, MT, Brazil}

\author{Bruno P. e Silva\orcidlink{0000-0001-5830-8903}}
\email{bruno.poti@fisica.ufc.br}
\affiliation{Department of Physics, University of Cear\'a, Fortaleza, CE, Brazil}
\affiliation{Federal Institute of Education, Science and Technology of Cear\'a - \textit{campus} Camocim, Camocim, CE, Brazil}

\author{Diego R. da Costa\orcidlink{0000-0002-1335-9552}} 
\email{diego_rabelo@fisica.ufc.br}
\affiliation{Department of Physics, University of Cear\'a, Fortaleza, CE, Brazil}
\affiliation{Department of Physics, University of Antwerp, Antwerp, Belgium}

\author{Alexandre C. Dias\orcidlink{0000-0001-5934-8528}} 
\email{alexandre.dias@unb.br}
\affiliation{Institute of Physics and International Center of Physics, University of Brasília, Brasília, DF, Brazil}

\author{Rodrigo G. Amorim\orcidlink{0000-0001-9611-8772}}
\email{rgamorim@id.uff.br}
\affiliation{Department of Physics -ICEx, Fluminense Federal University, Volta Redonda, RJ, 27213-145, Brazil}

\author{Teldo A. S. Pereira\orcidlink{0000-0001-7968-9194}} 
\email{teldo@fisica.ufmt.br}
\affiliation{Institute of Physics, University of Mato Grosso, Cuiab\'a, MT, Brazil}
\affiliation{National Institute of Science and Technology on Materials Informatics, Campinas, SP, Brazil}

\author{Jeanlex S. de Sousa\orcidlink{0000-0003-2836-0983}}
\email{jeanlex@fisica.ufc.br}
\affiliation{Department of Physics, University of Cear\'a, Fortaleza, CE, Brazil}

\begin{abstract}
Two-dimensional van der Waals heterostructures (vdWHs) based on transition metal dichalcogenides (TMDs) have emerged as promising platforms for next-generation low-power electronics and spintronics. Here, we investigate the structural, electronic, and spin properties of MoTe$_2$/HfS$_2$ vdWHs using first-principles density functional theory calculations. Our results reveal an intrinsic type--III (\textit{broken-gap}) band alignment, making this heterostructure highly attractive for tunnel field-effect transistor (TFET) applications. In addition, inversion-symmetry breaking at the interface, combined with strong spin--orbit coupling, induces a pronounced Rashba spin splitting around the $\Gamma$ point, accompanied by characteristic helical spin textures. We further show that external vertical electric fields, twist engineering, and both out-of-plane and in-plane strain provide efficient means to tune the band alignment, charge transfer, and Rashba parameters. In particular, electric fields and mechanical strain can drive reversible transitions between type--II and type--III band alignments while simultaneously modifying the Rashba spin texture and the band-to-band tunneling (BTBT) window. The twisted heterostructure with $\theta = 21.79^\circ$ exhibits similar tunability, including strain-induced transitions between distinct band-alignment regimes. These findings establish MoTe$_2$/HfS$_2$ vdWHs as versatile multifunctional platforms combining electrically tunable Rashba spin splitting with controllable type--II/type--III band alignment, offering promising opportunities for next-generation TFETs and spintronic devices.
\end{abstract}
\maketitle

\section{\label{sec_introduction}Introduction}

The relentless progress of modern electronics over the past decades has been largely driven by the continuous scaling of conventional metal--oxide--semiconductor field-effect transistors (MOSFETs). However, the further miniaturization of these devices faces fundamental physical limitations associated with power dissipation, thermal management, and the thermionic limit of the subthreshold swing (SS), which is restricted to approximately 60 mV/dec at room temperature due to the thermally activated carrier injection mechanism~\cite{Seabaugh2010,Ferain2011}. These challenges have stimulated the search for alternative transistor architectures capable of operating with higher energy efficiency and steeper switching characteristics. In this context, tunnel field-effect transistors (TFETs) have emerged as promising candidates. Their operation relies on band-to-band tunneling (BTBT), enabling subthreshold swings below the thermionic limit and opening a pathway toward low-power nanoelectronic devices~\cite{Sarkar2015,Ionescu2011,Lu6820751}.

Two-dimensional (2D) van der Waals heterostructures (vdWHs) based on transition metal dichalcogenides (TMDs) have attracted considerable attention owing to their remarkable electronic, optical, and spin-related properties, as well as their potential applications in nanoelectronics and spintronics~\cite{Rubstein2026,Lavor2021,Dias2021,CHENG2025177487}. The weak van der Waals interaction between adjacent layers allows the vertical stacking of different monolayers without stringent lattice-matching requirements, enabling the design of artificial heterostructures with tunable band alignments and interfacial properties~\cite{Mao2022,Lei2019,PhysRevMaterials.6.084001,ZHANG2024158986,C7CS00556C}. In particular, heterostructures exhibiting type--III (\textit{broken-gap}) band alignment are highly desirable for TFET applications because they facilitate efficient band-to-band tunneling between the valence-band states of one layer and the conduction-band states of the adjacent layer.~\cite{Iordanidou2024,Iordanidou2023,GUO2023155644,Lee2025BandAlignment}. Recent theoretical and experimental studies have demonstrated that type--III TMD-based vdWHs constitute promising platforms for low-power electronic and optoelectronic devices~\cite{Iordanidou2024,Iordanidou2023,Lee2025BandAlignment,Liang2025,Roy2016,Mitra2024,Roy2015MoS2WSe2}.

Type--III band alignment can be achieved through the stacking of selected TMD monolayers and related layered materials, such as MoTe$_2$, SnSe$_2$, ZrS$_2$, and HfS$_2$. Nevertheless, the number of experimentally realized and theoretically predicted 2D heterostructures that intrinsically exhibit a broken-gap alignment remains relatively limited, particularly those simultaneously combining favorable properties for TFET and spintronic applications. Consequently, the exploration of new material combinations capable of hosting both type--III band alignment and tunable electronic properties remains an important research direction. Motivated by this scenario, we focus here on the MoTe$_2$/HfS$_2$ van der Waals heterostructure.

Recent studies have shown that monolayer MoTe$_2$ exhibits temperature-dependent Rydberg excitons and trions, together with strong gate tunability, highlighting its potential for near-infrared optoelectronic applications~\cite{Biswas2023}. In addition, strain engineering in MoTe$_2$ has revealed significant opportunities for tailoring phase transitions, defect formation, and electronic properties, owing to the strong strain dependence of its structural phases~\cite{YANG2026438}. More recently, monolayer $T_d$--MoTe$_2$ has been reported to exhibit gate-controlled superconductivity~\cite{Rhodes2021MoTe2Superconductivity,Cui2019MoTe2SpinOrbit,PhysRevResearch.6.013132} and signatures of topological superconductivity~\cite{Li2024TopologicalMoTe2}.

On the other hand, monolayer HfS$_2$ possesses equally attractive electronic characteristics. Theoretical studies predict strain-induced indirect-to-direct band-gap transitions, broadening its potential for optoelectronic applications~\cite{HOAT2021165718}. Furthermore, significant improvements in its thermoelectric performance have been predicted under strain~\cite{Wang2021HfS2Thermoelectric}. Experimentally, few-layer HfS$_2$ transistors exhibit excellent current saturation and high on/off current ratios, demonstrating their suitability for ultra-low-power electronic devices~\cite{Kanazawa2016HfS2Transistors}.

Previous investigations of related heterostructures, such as ZrS$_2$/MoTe$_2$ and WTe$_2$/HfS$_2$, have shown that combining TMDs in the H and T phases can simultaneously induce type--III band alignment and Rashba spin splitting, opening new opportunities for multifunctional electronic devices, TFETs, and spintronic applications~\cite{Lei2019,Mao2022,Iordanidou2023}. Consequently, MoTe$_2$ and HfS$_2$ emerge as particularly promising building blocks for next-generation 2D heterostructures in which both charge and spin degrees of freedom can be simultaneously exploited and controlled.

The Rashba effect plays a fundamental role in the electronic states of surfaces and heterostructures, originating from the interplay between strong spin--orbit coupling (SOC) and inversion-symmetry breaking. This mechanism has become one of the cornerstones of modern spintronics because it enables electrical manipulation of spin polarization without the need for external magnetic fields~\cite{sharma2026,Bordoloi2024}. In vdWHs, interfacial charge redistribution and the resulting built-in electric field can break out-of-plane inversion symmetry, leading to momentum-dependent spin splitting near high-symmetry points of the Brillouin zone~\cite{Bordoloi2024,AFFANDI2019113611}. The resulting Rashba spin textures are highly attractive for spin manipulation, spin filtering, and spin transport applications in low-dimensional systems~\cite{sharma2026,Bordoloi2024,RevModPhys.76.323}. Moreover, the coexistence of Rashba spin splitting and type--III band alignment may enable the integration of spintronic and tunneling functionalities within a single material platform~\cite{Lei2019,Mao2022,Iordanidou2023,Bordoloi2024}.

Another appealing feature of vdWHs is the possibility of externally tuning their electronic properties through strain engineering, twist-angle modulation, and external electric fields~\cite{Iordanidou2023,Iordanidou2024,Peng2020,D3CP03591C,CastellanosGomez2013}. Recent studies have shown that vertical electric fields can substantially modify the relative positions of the valence- and conduction-band edges, inducing transitions between type--II and type--III band alignments and strongly affecting BTBT processes~\cite{Lei2019,GUO2023155644,Iordanidou2023,Iordanidou2024}. Likewise, twisted heterostructures can generate Moiré superlattices and alter the interlayer coupling, resulting in pronounced changes in the electronic structure, charge transfer, and correlated properties~\cite{D3CP03591C,wzz2-pszx}. These effects make electrically controlled and twisted vdWHs particularly attractive platforms for multifunctional nanoelectronic, spintronic, and quantum devices.

Motivated by these developments, we investigate the structural, electronic, and spin-related properties of the MoTe$_2$/HfS$_2$ van der Waals heterostructure using first-principles density functional theory (DFT) calculations including spin–orbit coupling and van der Waals corrections. Our results show that the energetically most stable stacking configuration lies remarkably close to the type-II/type-III band-alignment boundary. Furthermore, we explore the effects of external electric fields, mechanical strain, and twist-angle engineering, demonstrating that these external perturbations provide efficient control over charge transfer, band alignment, and spin properties, enabling reversible transitions between type-II and type-III band-alignment regimes. The computational details are provided in the Supplementary Information (SI) [40]. These findings establish the MoTe$_2$/HfS$_2$ heterostructure as a promising platform for TFETs, spintronic applications, and low-power multifunctional nanoelectronic devices. 

\section{RESULTS AND DISCUSSIONS}
\label{sec.results-and-discussions}
\subsection{Structural and Electronic Properties of  MoTe$_2$/HfS$_2$ Heterostructure}\label{sec.stability}
\begin{figure*}[!t]
\includegraphics[width=1\linewidth]{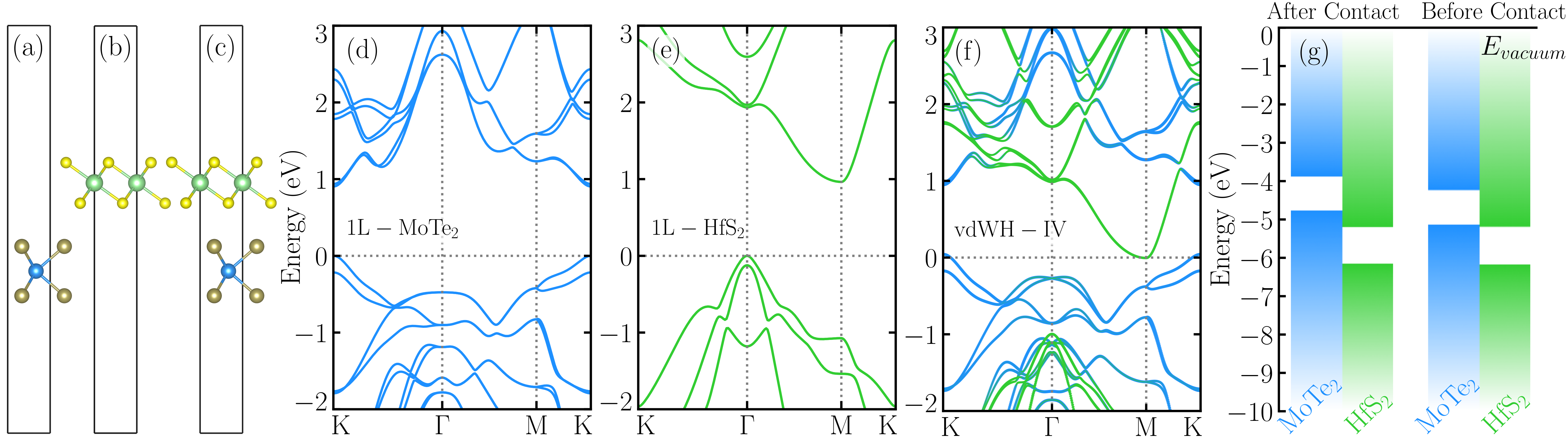}
\caption{(\textcolor{blue}{Color online}) (a)--(c) Side views of the crystal structures of 1L--MoTe$_2$, 1L--HfS$_2$, and the MoTe$_2$/HfS$_2$ heterostructure, respectively, illustrating the approach of the isolated monolayers leading to the formation of the van der Waals heterostructure (vdWH). The yellow, green, brown, and blue spheres represent S, Hf, Te, and Mo atoms, respectively. (d)--(e) Projected band structures of the isolated MoTe$_2$ (1L--MoTe$_2$) and HfS$_2$ (1L--HfS$_2$) monolayers, respectively. (f) Projected band structure of the MoTe$_2$/HfS$_2$ vdWH for the most stable stacking configuration, vdWH--IV, where the electronic contributions from MoTe$_2$ and HfS$_2$ are shown in blue and green, respectively. (g) Schematic band alignment diagrams with respect to the vacuum level for 1L--MoTe$_2$ and 1L--HfS$_2$ before and after vdWH formation, evidencing an intrinsic \textit{broken-gap} type--III band alignment.}
\label{fig:FIG1}
\end{figure*}

The MoTe$_2$/HfS$_2$ van der Waals heterostructure (vdWH) was constructed from the fully relaxed bulk phases of MoTe$_2$ and HfS$_2$. Bulk MoTe$_2$ was considered in its trigonal prismatic (hexagonal H) phase, with experimental lattice parameters of $a = 3.520$~\AA, $c = 13.910$~\AA, and unit-cell volume $V = 149.260$~\AA$^{3}$, as reported by Uesugi \textit{et al.}~\cite{Uesugi2017}. Bulk HfS$_2$ was modeled in its octahedral (T) phase, exhibiting experimental lattice parameters of $a = 3.623$~\AA, $c = 5.841$~\AA, and unit-cell volume $V = 66.398$~\AA$^{3}$, according to Rimmington $\&$ Balchin~\cite{Rimmington1974}.

To accurately describe the structural properties of the constituent materials, several exchange--correlation functionals were tested, including PBE~\cite{Perdew1996}, PBE$+$D2~\cite{Grimme2006,Kerber2008}, PBE$+$D3~\cite{Grimme2010}, PBEsol~\cite{Perdew2017}, vdW--DF2~\cite{Lee2010}, and rev-vdW--DF2~\cite{Hamada2014}. Among them, the PBE$+$D3 functional provided the best agreement with the experimental structural parameters of both bulk compounds (see Table S2 in Section I of the SI)~\cite{supplemental}) and was therefore adopted throughout this work.

Figures~\ref{fig:FIG1}(a) and \ref{fig:FIG1}(b) display the isolated MoTe$_2$ (1L--MoTe$_2$) and HfS$_2$ (1L--HfS$_2$) monolayers, respectively, while Fig.~\ref{fig:FIG1}(c) illustrates the resulting heterostructure. The yellow, green, brown, and blue spheres denote S, Hf, Te, and Mo atoms, respectively. Owing to the difference between the lattice parameters of the constituent monolayers, direct stacking would produce a lattice mismatch of approximately 3\%, which remains within the range experimentally achievable for van der Waals heterostructures. The lattice mismatch is defined as $\left[(a_{\mathrm{HfS}_{2}} - a_{\mathrm{MoTe}_{2}})/a_{\mathrm{MoTe}_{2}} \right]\times 100\%$. Under this condition, the HfS$_2$ layer would be subjected to compressive strain, whereas the MoTe$_2$ layer would experience tensile strain. To minimize this effect, the average lattice parameter obtained from the relaxed bulk structures was adopted for the heterostructure construction. This approach leads to a nearly uniform strain distribution between the layers and results in a residual strain of only $\sim1.5\%$. Considering the small magnitude of this deformation, the experimental realization of the proposed heterostructure is expected to be feasible.

We investigated six high-symmetry stacking configurations for the MoTe$_2$/HfS$_2$ vdWH, as illustrated in Figs.~S1(a)--(f) of the SI ~\cite{supplemental}. The considered configurations are: {\it i}. {\bf vdWH--I}, where the upper S atoms (S$_{up}$) are aligned with the Te atoms, while the Hf atoms occupy the Mo sites; {\it ii}. {\bf vdWH--II}, characterized by the alignment of Hf atoms with the Te sites and the lower S atoms (S$_{dw}$) with the Mo sites; {\it iii}. {\bf vdWH--III}, in which the S$_{dw}$ atoms are located above the Te atoms and the S$_{up}$ atoms above the Mo sites; {\it iv}. {\bf vdWH--IV}, where the S$_{up}$ atoms are aligned with the Te atoms, while the S$_{dw}$ atoms are positioned above the Mo atoms; {\it v}. {\bf vdWH--V}, in which the S$_{dw}$ atoms are located above the Te atoms and the Hf atoms above the Mo atoms; and {\it vi}. {\bf vdWH--VI}, where the Hf atoms are positioned above the Te atoms, while the S$_{up}$ atoms occupy the sites above the Mo atoms. The fully relaxed atomic structures corresponding to each stacking configuration are presented in Section~I of the SI~\cite{supplemental}.

To determine the most stable stacking arrangement, six high-symmetry configurations were considered, as shown in Figs.~S1(a)--(f) of the SI~\cite{supplemental}. The corresponding relaxed atomic structures are also provided in Section I of the SI. The interfacial stability was evaluated through the binding energy ($E_{b}$),
 
\[
E_b = (E_{\mathrm{MoTe}_2/\mathrm{HfS}_2} - E_{\mathrm{MoTe}_2} - E_{\mathrm{HfS}_2})/A,
\]
\begin{figure*}[!t]
\includegraphics[width=2.05\columnwidth]{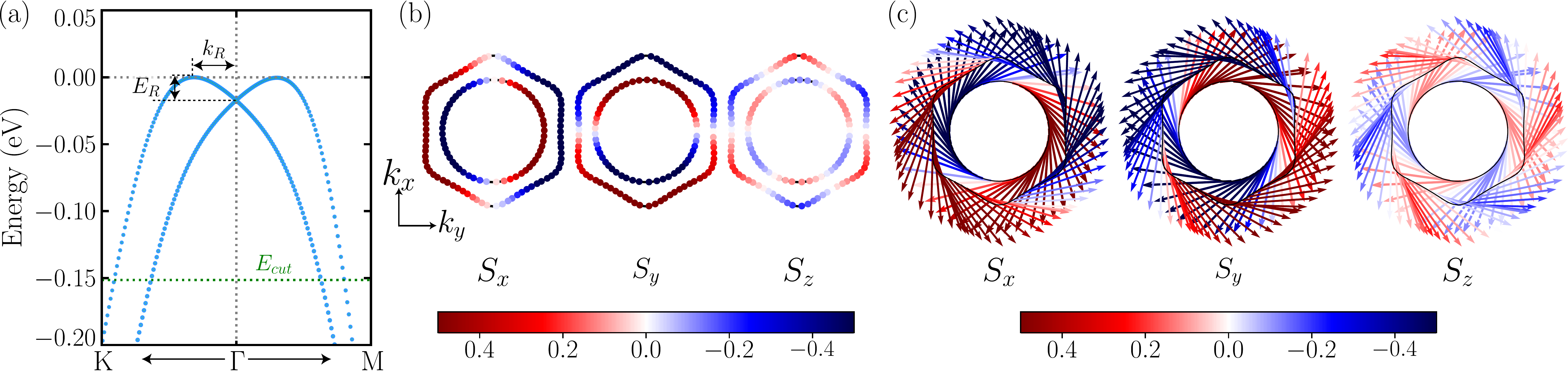}
\caption{(\textcolor{blue}{Color online}) (a) Rashba spin splitting around the $\Gamma$ point for the most stable MoTe$_2$/HfS$_2$ vdWH--IV heterostructure. The Rashba maximum momentum offset ($k_R$) and the energy difference between the band's extreme and the degenerate $\Gamma$ point of the band ($E_R$) are indicated by the black arrows, while the green dashed line marks the constant-energy cut ($E_{cut}$). (b) Spin polarization components $S_x$, $S_y$, and $S_z$ evaluated on the constant-energy contour in momentum space. The color scale represents the magnitude and sign of the spin polarization. (c) Corresponding spin textures in the $k_x$--$k_y$ plane, showing the Rashba-type helical spin-momentum locking induced by inversion symmetry breaking.}
\label{fig:FIG2}
\end{figure*}

\noindent where $E_{\mathrm{MoTe_2/HfS_2}}$ is the total energy of the heterostructure, $E_{\mathrm{MoTe_2}}$ and $E_{\mathrm{HfS_2}}$ are the energies of the strained isolated monolayers, and $A$ is the in-plane area of the unit cell. The calculated total energies, relative energies ($E-E_{\mathrm{vdWH-IV}}$), and binding energies are summarized in Table S3 of the SI~\cite{supplemental}. The analysis of both $E-E_{\mathrm{vdWH-IV}}$ and $E_b$ reveals that the vdWH--IV stacking configuration is the energetically most favorable structure, exhibiting a binding energy of $-26.430$~meV/\AA$^2$. This value falls within the range reported for other TMD-based van der Waals heterostructures~\cite{PhysRevMaterials.6.084001,Iordanidou2023,GUO2023155644,Iordanidou2024}. The negative binding energy confirms the stability of the heterostructure. The remaining stacking configurations also exhibit negative binding energies, ranging from $-18.649$ to $-26.430$~meV/\AA$^2$, indicating that all considered structures are energetically stable.

We also investigated the interlayer distance ($d_{\mathrm{int}}$) and the separation between the metallic Mo and Hf atoms ($d_{\mathrm{Mo-Hf}}$). For the vdWH--IV configuration, we obtained $d_{\mathrm{int}} = 3.158$~\AA\ and $d_{\mathrm{Mo-Hf}} = 6.427$~\AA. The calculated values of $d_{\mathrm{int}}$ are in good agreement with previously reported results~\cite{WANG2021148842}. Figures~S2(a)--(b) of the SI~\cite{supplemental} show the variation of $d_{\mathrm{Mo-Hf}}$ for the six high-symmetry stacking configurations considered in this work. Based on these results, all subsequent analyses were carried out using only the vdWH--IV configuration.

To further evaluate the thermal stability of the MoTe$_2$/HfS$_2$ vdWH, \textit{ab initio} molecular dynamics (AIMD) simulations were performed at 300~K for 5~ps. The corresponding results are presented in Fig.~S3 of the SI~\cite{supplemental}. As shown in Fig.~S3(a), the heterostructure remains structurally stable throughout the simulation, exhibiting neither structural reconstruction nor layer separation. Although a slight relative lateral displacement between the MoTe$_2$ and HfS$_2$ layers is observed after 5~ps, which can be attributed to the weak interlayer van der Waals interaction, the overall crystal framework is preserved. The top and side views of the final configuration, displayed as insets, further confirm the structural robustness of the heterostructure. Moreover, the total energy per atom exhibits only small fluctuations around its equilibrium value after the initial thermalization stage (approximately 0--1~ps), without any noticeable energy drift. Figure~S3(b) presents the temperature evolution during the AIMD simulation. Although moderate fluctuations are observed, particularly at the beginning of the trajectory, the temperature remains close to the target value of 300~K throughout the simulation. Such behavior is consistent with previous AIMD studies of thermally stable two-dimensional van der Waals heterostructures, including systems exhibiting type--III band alignment~\cite{Lei2019,Iordanidou2024}. These results demonstrate that the MoTe$_2$/HfS$_2$ vdWH is thermally stable at room temperature, supporting its potential experimental realization and its suitability for future nanoelectronic and spintronic device applications.

We now turn to the electronic properties of the constituent monolayers and the heterostructure. The electronic band structures were calculated within the PBE functional including spin--orbit coupling (PBE$+$SOC). Figure~\ref{fig:FIG1}(d) shows the band structure of 1L--MoTe$_2$, which exhibits a direct band gap of 0.90~eV at the K point. In contrast, Fig.~\ref{fig:FIG1}(e) displays the band structure of 1L--HfS$_2$, characterized by an indirect band gap of 0.96~eV. These values are slightly smaller than those previously reported in the literature~\cite{Zhang_2017,C9RA10087C}, but remain in good qualitative agreement. The reduction can be attributed to the small lattice strain ($\sim1.5\%$) introduced during the construction of the heterostructure, together with the inclusion of SOC effects. To verify this interpretation, we additionally calculated the band structures of the unstrained monolayers, shown in Figs.~S5(a)--(b) of the SI~\cite{supplemental}. In this case, the calculated band gaps are 1.01~eV for 1L--MoTe$_2$ and 1.18~eV for 1L--HfS$_2$, in closer agreement with previous theoretical reports~\cite{Zhang_2017,C9RA10087C}. We find that the applied strain does not alter the nature of the band gaps, but only induces a modest reduction in their magnitude. Therefore, the strained monolayers were adopted throughout this work to ensure consistency with the lattice-matched heterostructure calculations.

Figure~\ref{fig:FIG1}(f) presents the electronic band structure of the most stable MoTe$_2$/HfS$_2$ heterostructure (vdWH--IV). A qualitatively similar behavior is observed for the remaining stacking configurations, whose band structures are shown in Figs.~S6(a)--(f) of the SI~\cite{supplemental}. The electronic structure of the heterostructure can be understood, to first approximation, as a superposition of the bands of the isolated monolayers. The conduction-band minimum (CBM) is predominantly derived from the HfS$_2$ layer, whereas the valence-band maximum (VBM) originates mainly from the MoTe$_2$ layer, giving rise to a type--III (\textit{broken-gap}) band alignment. When SOC is included, the energy difference between the CBM and VBM is $\Delta E_{\mathrm{CBM-VBM}} = -0.048$~eV, indicating a finite overlap between the conduction- and valence-band edges. In contrast, a slightly positive value of $\Delta E_{\mathrm{CBM-VBM}} = 0.013$~eV is obtained without SOC. The band structures calculated with and without SOC for the remaining stacking configurations, together with the corresponding values of $\Delta E_{\mathrm{CBM-VBM}}$, are provided in Figs.~S6 and S7 and Table~S3 of the SI~\cite{supplemental}, respectively.
\begin{figure*}[!t]
\includegraphics[width=1\linewidth]{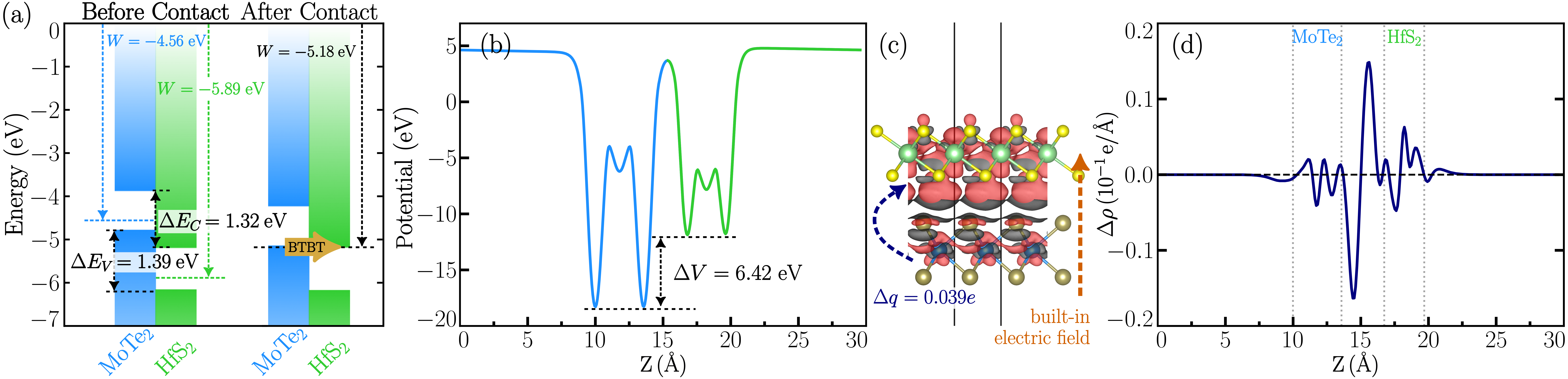}
\caption{(\textcolor{blue}{Color online}) (a) Schematic band alignment of the MoTe$_2$/HfS$_2$ heterostructure before and after contact, referenced to the vacuum level. The work functions, conduction-band offset ($\Delta E_{\mathrm{CBM}}$), and valence-band offset ($\Delta E_{\mathrm{VBM}}$) are indicated, evidencing a broken-gap type--III band alignment and the possibility of band-to-band tunneling (BTBT). (b) Planar-averaged electrostatic potential along the out-of-plane direction ($z$) for the MoTe$_2$/HfS$_2$ vdWH, showing the potential difference ($\Delta V$ ) between the layers. (c) Three-dimensional charge-density-difference isosurface of the heterostructure, illustrating charge transfer from MoTe$_2$ to HfS$_2$ and the resulting built-in electric field. Red and black isosurfaces represent charge accumulation and depletion, respectively. (d) Plane-averaged charge-density difference $\Delta \rho(z)$ along the $z$ direction, highlighting the interfacial charge redistribution between the MoTe$_2$ and HfS$_2$ layers.}
\label{fig:FIG3}
\end{figure*}

To overcome the well-known band-gap underestimation of semilocal functionals, additional calculations were performed using the hybrid HSE06 functional~\cite{Krukau2006}. Without SOC, HSE06 predicts a larger positive value of $\Delta E_{\mathrm{CBM-VBM}}=0.23$ eV, corresponding to a type-II band alignment, in agreement with previous theoretical studies~\cite{WANG2021148842}. When SOC is included, however, the energy separation is substantially reduced to $\Delta E_{\mathrm{CBM-VBM}}=0.057$ eV, placing the system very close to the type--II/type--III transition. Therefore, although the exact value of $\Delta E_{\mathrm{CBM-VBM}}$ depends on the exchange--correlation functional, both PBE and HSE06 consistently show that spin--orbit coupling strongly reduces the energy separation between the conduction- and valence-band edges, driving the MoTe$_2$/HfS$_2$ heterostructure toward the \textit{broken-gap} regime.

Since the semilocal PBE$+$SOC calculations predict an intrinsic \textit{broken-gap} alignment while HSE06+SOC places the system only 57 meV away from this regime, the two approaches lead to the same qualitative physical picture: the MoTe$_2$/HfS$_2$ heterostructure lies in the immediate vicinity of the type--II/type--III boundary. Consequently, external perturbations such as electric fields, strain, or twist angle are expected to readily tune the system across this boundary, justifying the use of the computationally more efficient PBE$+$SOC approach to investigate these trends.

The out-of-plane inversion-symmetry breaking induced in 1L--MoTe$_2$ by its stacking with 1L--HfS$_2$ gives rise to Rashba-type spin splitting in the vicinity of the $\Gamma$ point, where bands associated with opposite spin states are shifted in opposite directions in momentum space. This behavior is analogous to that observed in heavy-metal surfaces~\cite{PhysRevLett.77.3419,Koroteev2004,PhysRevLett.98.186807}. Notably, such spin splitting is absent in the isolated MoTe$_2$ monolayer and emerges only after the formation of the MoTe$_2$/HfS$_2$ vdWH, demonstrating that the Rashba effect originates from the interfacial inversion-symmetry breaking combined with strong spin--orbit coupling.

Figure~\ref{fig:FIG2}(a) presents an enlarged view of the electronic band structure around the $\Gamma$ point, highlighting the Rashba spin splitting along the K--$\Gamma$ high-symmetry path. The characteristic Rashba parameters, namely the momentum offset ($k_R$), the Rashba energy ($E_R$), defined as the energy difference between the band extremum and the spin-degenerate point at $\Gamma$, and the Rashba coefficient ($\alpha_R = 2E_R/k_R$), are also indicated. The calculated values are $k_R = 0.136$~\AA$^{-1}$, $E_R = 17.256$~meV, and $\alpha_R = 0.254$~eV$\cdot$\AA. These values lie within the range reported for TMD-based vdW heterostructures~\cite{Iordanidou2023,Iordanidou2024}, are comparable to those observed for Au(111) surfaces~\cite{PhysRevLett.77.3419}, and are significantly larger than those reported for the InGaAs/InAlAs quantum well~\cite{PhysRevLett.78.1335}. Enlarged views of the Rashba-split bands around the $\Gamma$ point for the remaining stacking configurations, together with the corresponding values of $k_R$, $E_R$, and $\alpha_R$, are presented in Figs.~S10(a)--(f) and Table~S3 of the Supporting Information (SI)~\cite{supplemental}, respectively.

Figure~\ref{fig:FIG2}(b) shows the spin-polarization components $S_x$, $S_y$, and $S_z$ calculated on a constant-energy contour ($E_{cut}$), defined by the green horizontal line in Fig.~\ref{fig:FIG2}(a), while Fig.~\ref{fig:FIG2}(c) displays the corresponding spin texture in reciprocal space. The in-plane components, $S_x$ and $S_y$, exhibit a characteristic helical spin arrangement, whereas the color scale indicates the degree of spin polarization. The spin polarization is considerably stronger within the plane ($x,y$), where two approximately circular contours can be identified around the $\Gamma$ point. These inner and outer contours exhibit opposite spin helicities, with clockwise and counterclockwise spin rotations, respectively. They correspond to the Rashba-split bands and provide clear evidence of spin--momentum locking. In contrast, the out-of-plane component, $S_z$, shows a non-zero but considerably weaker signal than the in-plane components. This residual out-of-plane spin polarization can be attributed to the crystal symmetry of the heterostructure and the strong spin--orbit coupling primarily associated with the MoTe$_2$ layer. These results suggest that the MoTe$_2$/HfS$_2$ vdWH is a promising platform for spintronic devices based on electrical control of spin degrees of freedom.

It is worth noting that type--III vdWHs may exhibit negative differential resistance (NDR) arising from BTBT processes~\cite{Xia2018,Lei2019,Ferdous2024,Mitra2024,Liang2025}. To further elucidate the physical mechanism associated with the broken-gap band alignment, we constructed the schematic representation shown in Fig.~\ref{fig:FIG3}(a).

The work function, which serves as an intrinsic reference for band alignment, is defined as $W = E_{vac} - E_{F}$, where $E_{vac}$ and $E_{F}$ denote the vacuum level and Fermi level, respectively. As shown in Fig.~\ref{fig:FIG3}(a), the conduction-band offset between the HfS$_2$ and MoTe$_2$ monolayers is $\Delta E_{C} = 1.32$~eV, while the corresponding valence-band offset is $\Delta E_{V} = 1.39$~eV. Before heterostructure formation, the calculated work functions of 1L--MoTe$_2$ and 1L--HfS$_2$ are 4.56 and 5.89~eV, respectively.

Upon formation of the heterostructure, spontaneous charge transfer occurs from the MoTe$_2$ layer to the HfS$_2$ layer owing to the difference in their work functions, giving rise to an electrostatic potential difference of $\Delta V = 6.42$~eV, as illustrated in Fig.~\ref{fig:FIG3}(b). As a consequence, the Fermi level of the heterostructure shifts to a position located below the CBM of HfS$_2$ and above the VBM of MoTe$_2$. The resulting work function of the vdWH is 5.18~eV, which is higher than that of isolated MoTe$_2$ and slightly lower than that of isolated HfS$_2$. Consequently, the heterostructure exhibits a type--III (\textit{broken-gap}) band alignment, enabling efficient electron tunneling from the VBM of MoTe$_2$ to the CBM of HfS$_2$, which is the hallmark of the BTBT mechanism. This characteristic makes the MoTe$_2$/HfS$_2$ heterostructure particularly attractive for TFETs applications.

To gain deeper insight into the interlayer coupling mechanism and charge redistribution in the MoTe$_2$/HfS$_2$ vdWH, we calculated both the three-dimensional (3D) charge-density-difference isosurface [Fig.~\ref{fig:FIG3}(c)] and the planar-averaged charge-density difference along the $c$ direction [Fig.~\ref{fig:FIG3}(d)]. The planar-averaged charge-density difference is defined as

\begin{equation}
 \Delta \rho (z) = \Delta\rho_{\mathrm{MoTe}_{2}/\mathrm{HfS}_{2}} - \Delta\rho_{\mathrm{HfS}_{2}} - \Delta\rho_{\mathrm{MoTe}_{2}},    
\end{equation}

\noindent where $\Delta\rho_{\mathrm{MoTe}_{2}/\mathrm{HfS}_{2}}$ denotes the charge density of the MoTe$_2$/HfS$_2$ heterostructure, while $\Delta\rho_{\mathrm{HfS}_{2}}$ and $\Delta\rho_{\mathrm{MoTe}_{2}}$ correspond to the charge densities of the isolated 1L--HfS$_2$ and 1L--MoTe$_2$ monolayers, respectively.
\begin{figure}[!t]
\includegraphics[width=1\columnwidth]{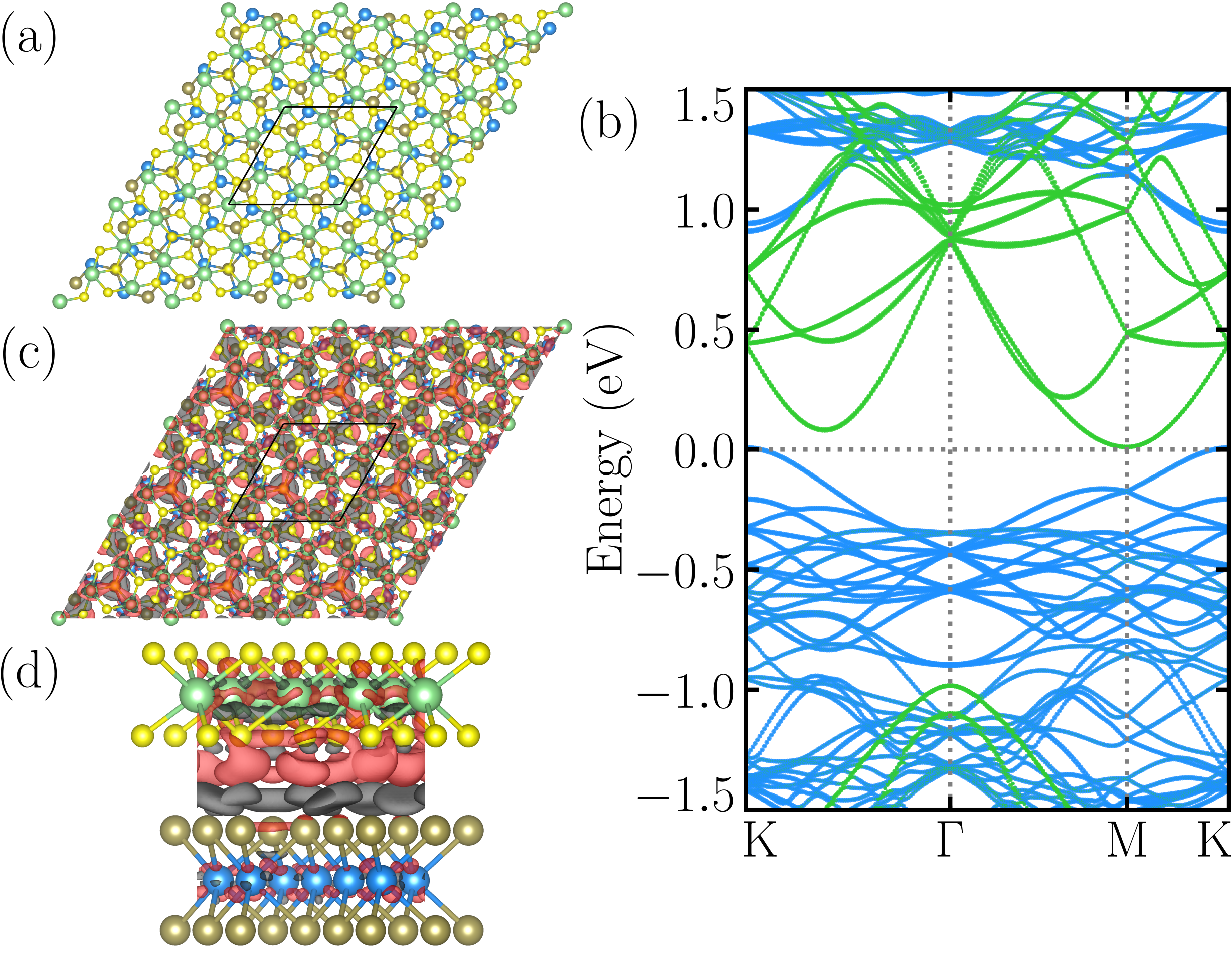}
\caption{(\textcolor{blue}{Color online}) (a) Top view of the twisted MoTe$_2$/HfS$_2$ heterostructure (t-MoTe$_2$/HfS$_2$) with a rotation angle of $\theta = 21.79^\circ$, highlighting the moiré supercell (black rhombus). (b) Electronic band structure of the twisted system, calculated using the PBE$+$SOC, showing contributions from MoTe$_2$ (blue) and HfS$_2$ (green). (c)--(d) Three-dimensional charge density difference of the twisted heterostructure, illustrating the spatial modulation induced by the moiré pattern, where red and black isosurfaces indicate charge accumulation and depletion, respectively. The isosurface value is set to $2.5 \times 10^{-4}$ $e$\AA$^{-3}$.}
\label{fig:FIG4}
\end{figure}

The 3D charge-density-difference isosurface shown in Fig.~\ref{fig:FIG3}(c) reveals a net charge transfer from the MoTe$_2$ layer to the HfS$_2$ layer. The red and black regions indicate electron accumulation and depletion, respectively. Consistently, the planar-averaged charge-density difference displayed in Fig.~\ref{fig:FIG3}(d) shows that the charge redistribution occurs predominantly between the Te atoms of the MoTe$_2$ layer and the adjacent S atoms of the HfS$_2$ layer. 

To quantitatively evaluate this process, Bader charge analysis was performed. The calculated charge transfer associated with each atomic species is Hf($-2.016e$), S($+2.055e$), Mo($-0.490e$), and Te($+0.451e$), where positive and negative values denote electron gain and loss, respectively. As a result, a net charge transfer of $\Delta q = 0.039e$ per unit cell occurs from the MoTe$_2$ layer to the HfS$_2$ layer, a value comparable to those reported for other TMD-based van der Waals heterostructures~\cite{Lei2019,Mao2022,GUO2023155644,Liang2025}. This interfacial charge redistribution allows the MoTe$_2$/HfS$_2$ vdWH to be viewed as a nanoscale parallel-plate capacitor, generating an intrinsic electric field directed from the MoTe$_2$ layer toward the HfS$_2$ layer.

We now turn our attention to the twisted MoTe$_2$/HfS$_2$ heterostructure ($t$--MoTe$_2$/HfS$_2$). Twisted TMD-based systems have attracted considerable interest in recent years owing to their rich electronic properties and the emergence of Moiré-driven phenomena~\cite{Rosenberger2020,Tran_2021,Devakul2021,Zhao2021_phosphorene_moire}. Moreover, recent advances in fabrication techniques have made the experimental realization of twisted van der Waals devices increasingly feasible. Motivated by these developments, we next investigate the structural and electronic properties of the $t$--MoTe$_2$/HfS$_2$ heterostructure.
\begin{figure*}[!t]
\includegraphics[width=1\linewidth]{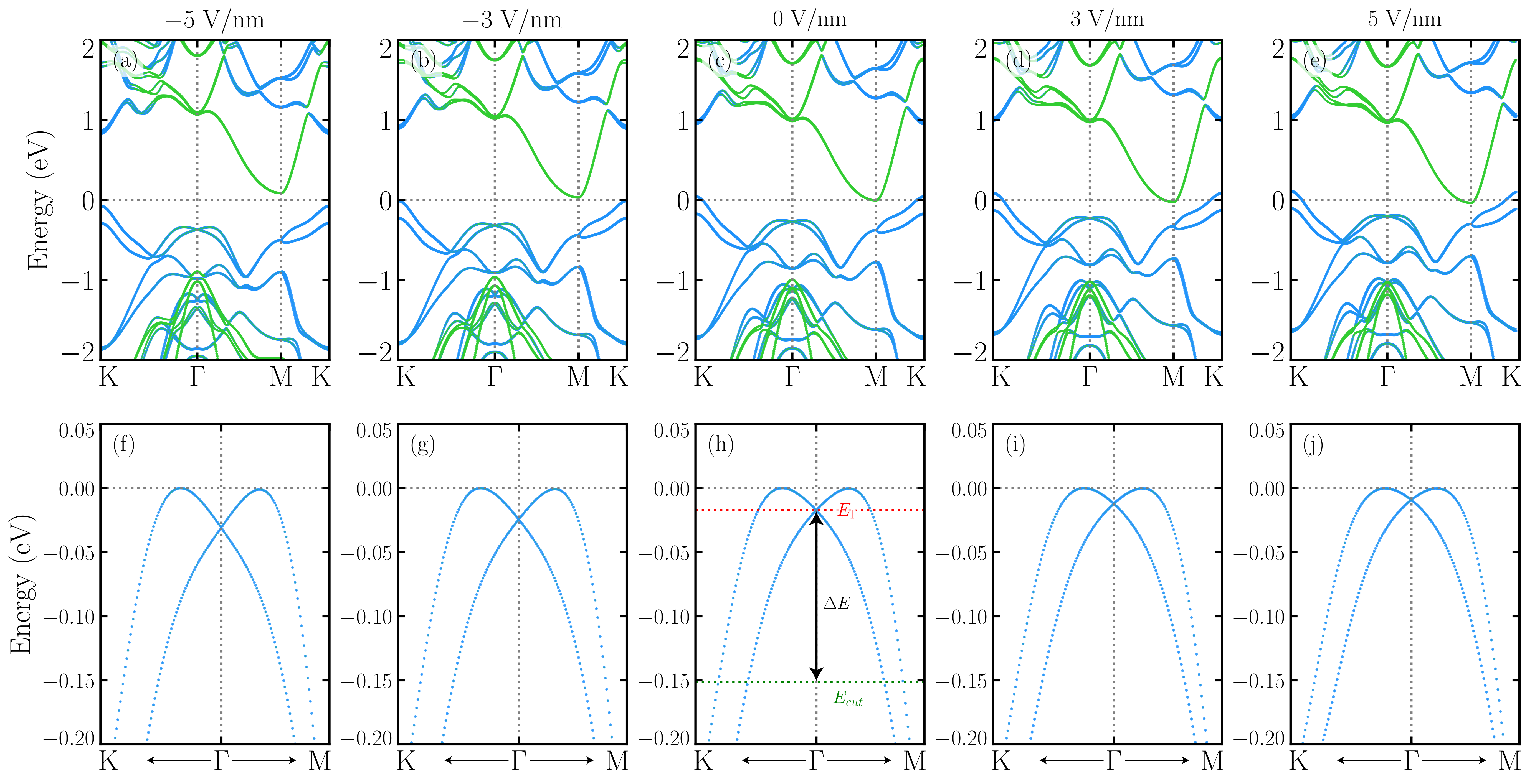}
\caption{(\textcolor{blue}{Color online}) (a)--(e) Electronic band structures of the MoTe$_2$/HfS$_2$ vdWH under external electric fields ranging from $-5$ to $5$~V/nm, calculated using the PBE$+$SOC approach. Panels (f)--(j) show an enlarged view around the $\Gamma$ point, highlighting the Rashba spin splitting. The red dashed line denotes the energy of the Rashba degeneracy point in $\Gamma$ ($E_{\Gamma}$), while the green dotted line indicates a cutoff energy ($E_{cut}$) and $\Delta E$ is a small energy shift relative to $E_{\Gamma}$.}
\label{fig:FIG5}
\end{figure*}

Four twisted MoTe$_2$/HfS$_2$ heterostructures were constructed from the equilibrium vdWH by employing the coincidence-site lattice~\cite{Koda2016, Uchida2014} approach described above. For each commensurate supercell, the twist angle is uniquely determined by an integer pair $(m,n)$, which defines the coincidence lattice vectors and, consequently, the size of the Moiré supercell. Owing to the hexagonal symmetry of the heterostructure, several commensurate Moiré superlattices can be generated following this construction, as previously demonstrated for twisted boron arsenide bilayers ($t$--BAs) and other van der Waals materials~\cite{Lopes2026,Uchida2014,D3CP03591C,wzz2-pszx}. To ensure computational feasibility, only supercells containing fewer than 150 atoms ($N_{\mathrm{atom}}<150$) were considered. The selected commensurate structures correspond to the integer pairs $(m,n)=(2,3)$, $(1,2)$, $(1,3)$, and $(1,4)$, yielding twist angles of $\theta=13.17^\circ$, $21.79^\circ$, $32.20^\circ$, and $38.21^\circ$, respectively. These structures contain 114, 42, 78, and 126 atoms per supercell.

Figure~\ref{fig:FIG4}(a) shows the Moiré superlattice corresponding to $\theta = 21.79^\circ$, where the black rhombus highlights the supercell. The remaining twisted structures are presented in Fig.~S4 of the SI~\cite{supplemental}. Full structural relaxations, including both atomic positions and lattice parameters, were performed for all twisted heterostructures. The resulting interlayer distances are 3.359~\AA\ ($\theta = 13.17^\circ$), 3.387~\AA\ ($\theta = 21.79^\circ$), 3.415~\AA\ ($\theta = 32.20^\circ$), and 3.403~\AA\ ($\theta = 38.21^\circ$). These values are slightly larger than that obtained for the untwisted heterostructure, reflecting the influence of Moiré-induced structural relaxation, in agreement with previous reports on twisted van der Waals systems~\cite{Lopes2026,Uchida2014,wzz2-pszx}. 

The electronic band structures of the twisted MoTe$_2$/HfS$_2$ heterostructures calculated within the PBE$+$SOC framework are presented in Fig.~S11 of the Supplementary Information (SI)~\cite{supplemental}. For all twist angles, the orbital-projected band structures reveal that the conduction-band minimum (CBM) remains predominantly derived from the HfS$_2$ layer (green), whereas the valence-band maximum (VBM) is mainly contributed by the MoTe$_2$ layer (blue). Among the investigated structures, the twist angles $\theta=21.79^\circ$ [Fig.~\ref{fig:FIG4}(b)] and $\theta=32.20^\circ$ [Fig.~S11(c) of the SI~\cite{supplemental}] exhibit the smallest energy separations between the CBM and VBM, with $\Delta E_{\mathrm{CBM-VBM}}=4$ and 12~meV, respectively, placing these systems at the verge of the transition between the type-II and type-III (broken-gap) band-alignment regimes. In contrast, the twisted structures with $\theta=13.17^\circ$ and $38.21^\circ$ [Figs.~S11(a) and S11(d) of the SI~\cite{supplemental}] display larger negative values of $\Delta E_{\mathrm{CBM-VBM}}$, corresponding to a more pronounced type-III band alignment. These results demonstrate that twist-angle engineering provides an effective strategy for continuously tuning the band-edge alignment while preserving the spatial separation of electrons and holes across the MoTe$_2$/HfS$_2$ interface.

Figures~\ref{fig:FIG4}(c)--(d) present the charge-density difference of the twisted heterostructure, where the red and black regions indicate electron accumulation and depletion, respectively. Similar to the untwisted MoTe$_2$/HfS$_2$ vdWH, charge transfer occurs from the MoTe$_2$ layer to the HfS$_2$ layer. Bader charge analysis reveals a net charge transfer of approximately $\Delta q = 0.236e$ per supercell, substantially larger than that obtained for the untwisted heterostructure. Owing to the high computational cost associated with twisted Moiré superlattices, the effects of external electric fields as well as out-of-plane and in-plane strain are investigated only for the $\theta = 21.79^\circ$ configuration in the following sections.

\subsection{Electronic properties under an applied vertical electric field}\label{sec.electric-field}

Gate-voltage modulation is one of the most widely employed approaches in experimental studies and practical electronic devices, providing an effective route for tuning the electronic properties of low-dimensional materials~\cite{Radisavljevic2011,Deng2017,D3CP03591C,Lopes2025,Xu2026}. Such an approach enables the controlled manipulation of electronic characteristics to meet the requirements of different technological applications. Motivated by this capability, we investigate the effects of an external vertical electric field ($E_{\mathrm{ext}}$) on the electronic properties of both the MoTe$_2$/HfS$_2$ vdWH and the twisted heterostructure $t$--MoTe$_2$/HfS$_2$.
\begin{figure}[!t]
\includegraphics[width=0.75\columnwidth]{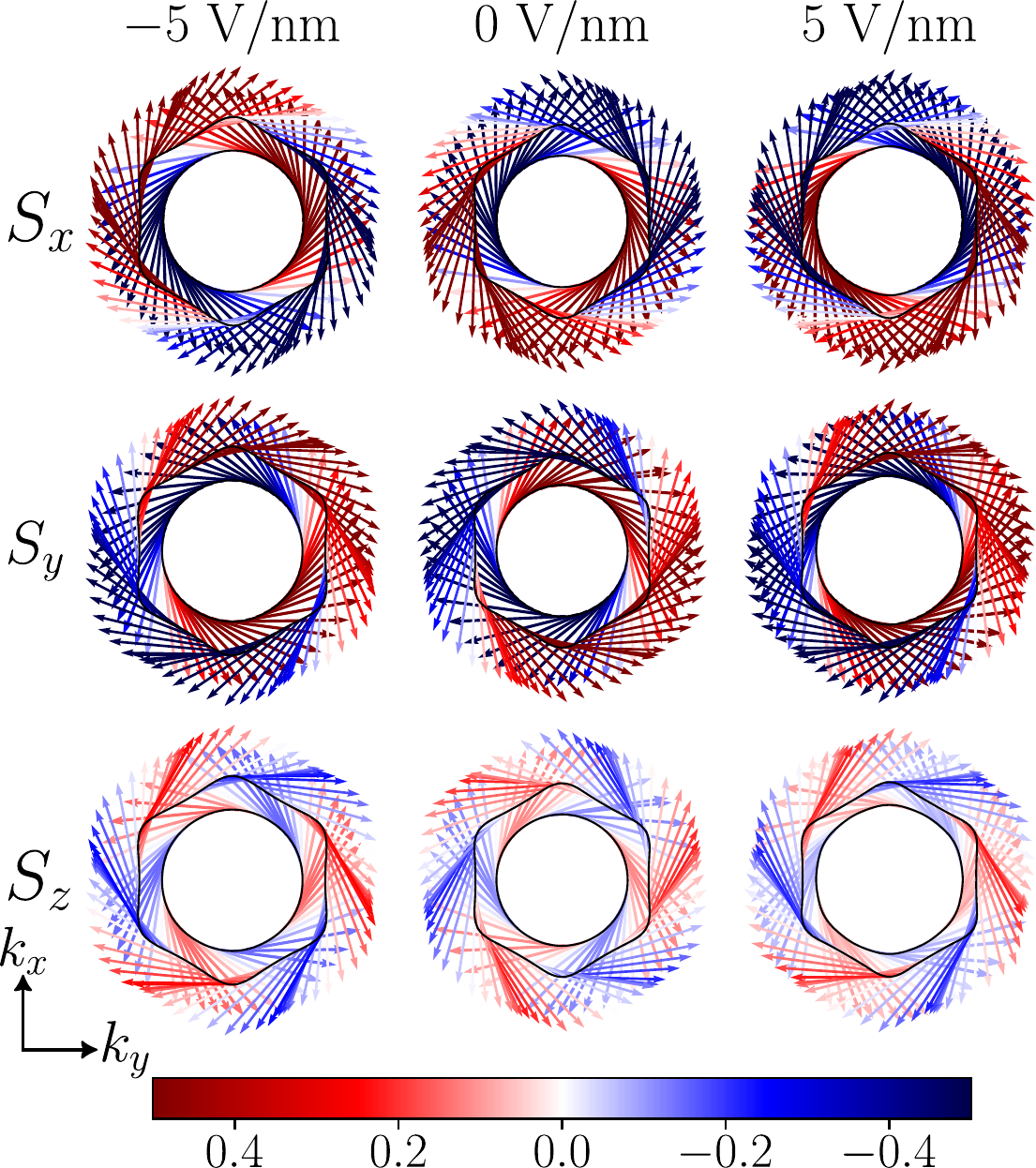}
\caption{(\textcolor{blue}{Color online}) Spin textures of the MoTe$_2$/HfS$_2$ vdWH under external electric fields of $-5$, $0$, and $5$~V/nm. The first, second, and third rows correspond to the spin projections $S_x$, $S_y$, and $S_z$, respectively. The arrows indicate the in-plane and out-of-plane spin components, while the color scale represents the magnitude and sign of the spin polarization, around the $\Gamma$ point.}
\label{fig:FIG6}
\end{figure}

The external electric field is applied perpendicular to the stacked layers and was included in both the structural relaxation and electronic-structure calculations, following methodologies previously reported in the literature~\cite{PhysRevMaterials.6.084001,Iordanidou2023,Iordanidou2024,Lopes2025}. A positive electric field ($E_{\mathrm{ext}} > 0$) is defined in the direction from the MoTe$_{2}$ layer to the HfS$_{2}$ layer, while a negative field ($E_{\mathrm{ext}} < 0$) is applied in the opposite direction, from HfS$_{2}$ to MoTe$_{2}$. The positive $E_{ext}$ has the same direction as the internal electric field, generated by the interfacial charge transfer between the layers, and thus the internal electric field is intensified. In contrast, for $E_{\mathrm{ext}} < 0$, the internal electric field is partially attenuated. The external electric field values considered ranged from $-5$~V/nm to $5$~V/nm.

Figures~\ref{fig:FIG5}(a)--(e) display the electronic band structures of the MoTe$_2$/HfS$_2$ vdWH under external vertical electric fields, calculated within the PBE$+$SOC framework. For positive electric fields [Figs.~\ref{fig:FIG5}(d)--(e)], the CBM associated with the HfS$_2$ layer (green bands) shifts toward lower energies relative to the Fermi level, whereas the VBM derived from the MoTe$_2$ layer (blue bands) moves toward higher energies. As a consequence, the energy difference between the CBM and VBM decreases from $-0.11$~eV at $E_{\mathrm{ext}} = 3$~V/nm to $-0.15$~eV at $E_{\mathrm{ext}} = 5$~V/nm. The increasingly negative values of $\Delta E_{\mathrm{CBM-VBM}}$ indicate an enhanced overlap between the valence- and conduction-band edges, thereby strengthening the BTBT process. Consequently, a larger number of carriers are expected to tunnel from the VBM of MoTe$_2$ into the CBM of HfS$_2$. Therefore, under positive electric fields, the type--III (\textit{broken-gap}) band alignment is not only preserved but further reinforced as the field strength increases.
\begin{figure}[!b]
\includegraphics[width=1\columnwidth]{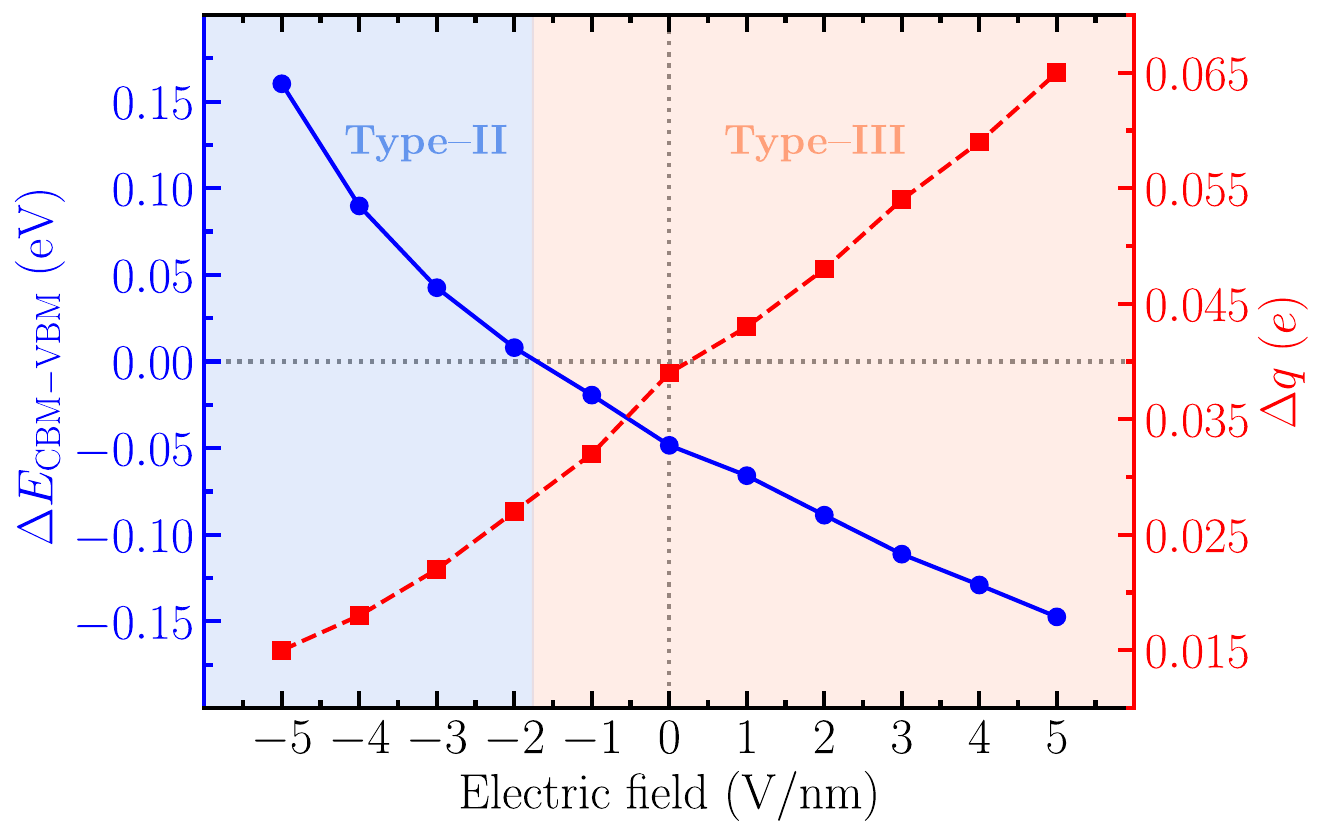}
\caption{(\textcolor{blue}{Color online}) Electric-field dependence of the band alignment and charge transfer in the MoTe$_2$/HfS$_2$ vdWH. The blue curve (left axis) shows the energy difference $\Delta E_{\mathrm{CBM-VBM}}$, while the red curve (right axis) represents the Bader charge transfer $\Delta q$. The shaded regions indicate the transition between type--II (light blue) and type--III (light orange) band alignment. $\Delta q$ is charge transfer from the MoTe$_2$ layer to the HfS$_2$ layer.}
\label{fig:FIG7}
\end{figure}

\begin{figure*}[!t]
\includegraphics[width=0.95\linewidth]{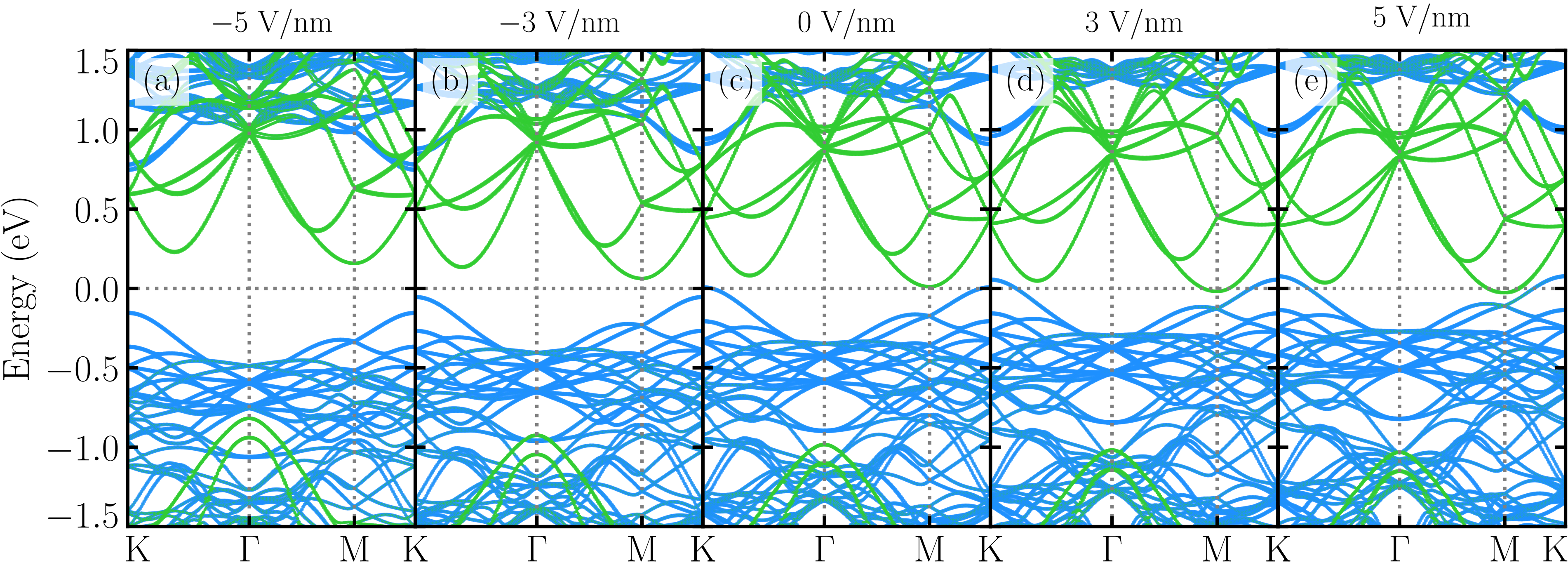}
\caption{(\textcolor{blue}{Color online}) Electronic band structures of the twisted $t$--MoTe$_2$/HfS$_2$ vdWH calculated using the PBE$+$SOC approach under different external vertical electric fields, ranging from $-5$ to $5$ V/nm. The blue and green colors represent the contributions from the MoTe$_2$ and HfS$_2$ layers, respectively.}
\label{fig:FIG8}
\end{figure*}

In contrast, an opposite trend is observed for $E_{\mathrm{ext}} < 0$. For $E_{\mathrm{ext}} = -3$ and $-5$~V/nm, the energy differences between the CBM and VBM are $0.043$ and $0.16$~eV, respectively, indicating the opening of a finite band gap that increases with the magnitude of the applied electric field. Under these conditions, the heterostructure undergoes a transition from a type--III (\textit{broken-gap}) to a type--II band alignment. These findings are consistent with previous reports on TMD-based vdWHs exhibiting type--III band alignment~\cite{Lei2019,Mao2022,Iordanidou2023,GUO2023155644}. The shift of the band edges can be attributed to the electrostatic potential gradient across the interface induced by the external electric field, which promotes charge redistribution between the two layers. The equilibrium interlayer distance of the most stable heterostructure (vdWH--IV) is approximately $0.32$~nm. Therefore, considering the maximum applied field of $\pm5$~V/nm, the corresponding potential difference between the layers is approximately $\Delta V = \pm1.6$~V. Although this value is relatively large, perpendicular electric fields of comparable or even greater magnitude have already been experimentally achieved in two-dimensional heterostructures through dual ionic gating techniques~\cite{Chu2015,Weintrub2022}. These experimental demonstrations indicate that the electric-field regime investigated here lies within experimentally accessible conditions for vdWH-based devices.

Figures~\ref{fig:FIG5}(f)--(j) illustrate the evolution of the Rashba-split bands in the vicinity of the $\Gamma$ point under different external electric fields. The perpendicular electric field strongly affects both the Rashba energy ($E_R$) and the momentum offset ($k_R$). For $E_{\mathrm{ext}} < 0$, the Rashba splitting becomes progressively more pronounced, whereas it is gradually suppressed for $E_{\mathrm{ext}} > 0$.

The red dashed horizontal line denotes the energy of the Rashba degeneracy point at $\Gamma$ ($E_{\Gamma}$), while the green dashed line corresponds to a constant-energy contour defined as $E_{cut}=E_{\Gamma}+\Delta E$, where $\Delta E=134.225$~meV is kept fixed for all electric-field values. Besides shifting the Rashba degeneracy point in energy, the external electric field also modifies the curvature and separation of the Rashba-split bands. For $E_{\mathrm{ext}}=-5$~V/nm, the Rashba parameters reach $E_R = 30.793$~meV, $k_R = 0.161$~\AA$^{-1}$ and $\alpha_R = 0.382$~eV$\cdot$\AA, indicating a substantial enhancement of the Rashba effect compared with the field-free case.

To gain further insight into the microscopic origin of this behavior, Fig.~\ref{fig:FIG6} presents the spin textures of the MoTe$_2$/HfS$_2$ vdWH under different external electric fields. The first and second rows correspond to the in-plane spin components ($S_x$ and $S_y$), whereas the third row displays the out-of-plane component ($S_z$). A clear modification of the spin texture is observed upon reversing the direction of the applied electric field. For $E_{\mathrm{ext}}=-5$~V/nm, the $S_x$ component exhibits a well-defined helical spin rotation. When the field is reversed to $E_{\mathrm{ext}}=+5$~V/nm, the helicity is preserved, while the spin polarization is inverted, as evidenced by the interchange of the red and blue regions. In contrast, the $S_y$ and $S_z$ components display similar rotational patterns and spin polarizations for both field directions. These results demonstrate that the external electric field enables an efficient modulation of the Rashba spin texture. Furthermore, the spin polarization is slightly stronger for $E_{\mathrm{ext}}<0$ than for $E_{\mathrm{ext}}>0$, in agreement with the larger Rashba splitting observed under negative electric fields.

Figure~\ref{fig:FIG7} summarizes the evolution of the band alignment and charge transfer in the MoTe$_2$/HfS$_2$ vdWH as a function of the external electric field. The blue curve represents the energy difference between the HfS$_2$-derived CBM and the MoTe$_2$-derived VBM, $\Delta E_{\mathrm{CBM-VBM}}$, while the red curve corresponds to the Bader charge transfer ($\Delta q$) between the two layers.

A nearly linear variation of $\Delta E_{\mathrm{CBM-VBM}}$ is observed with increasing electric field. For negative fields, $\Delta E_{\mathrm{CBM-VBM}}$ remains positive, indicating a type--II band alignment. In this regime, the electronic states remain energetically separated, reducing the overlap between the valence- and conduction-band edges and consequently suppressing interlayer tunneling. As the electric field increases toward positive values, $\Delta E_{\mathrm{CBM-VBM}}$ decreases continuously and becomes negative at approximately $E_{\mathrm{ext}}\approx-1.8$~V/nm. This behavior signals a gradual transition from a type--II to a type--III (\textit{broken-gap}) band alignment, enhancing the overlap between the MoTe$_2$-derived VBM and the HfS$_2$-derived CBM and thereby favoring band-to-band tunneling processes.
\begin{figure}[!t]
\includegraphics[width=1\columnwidth]{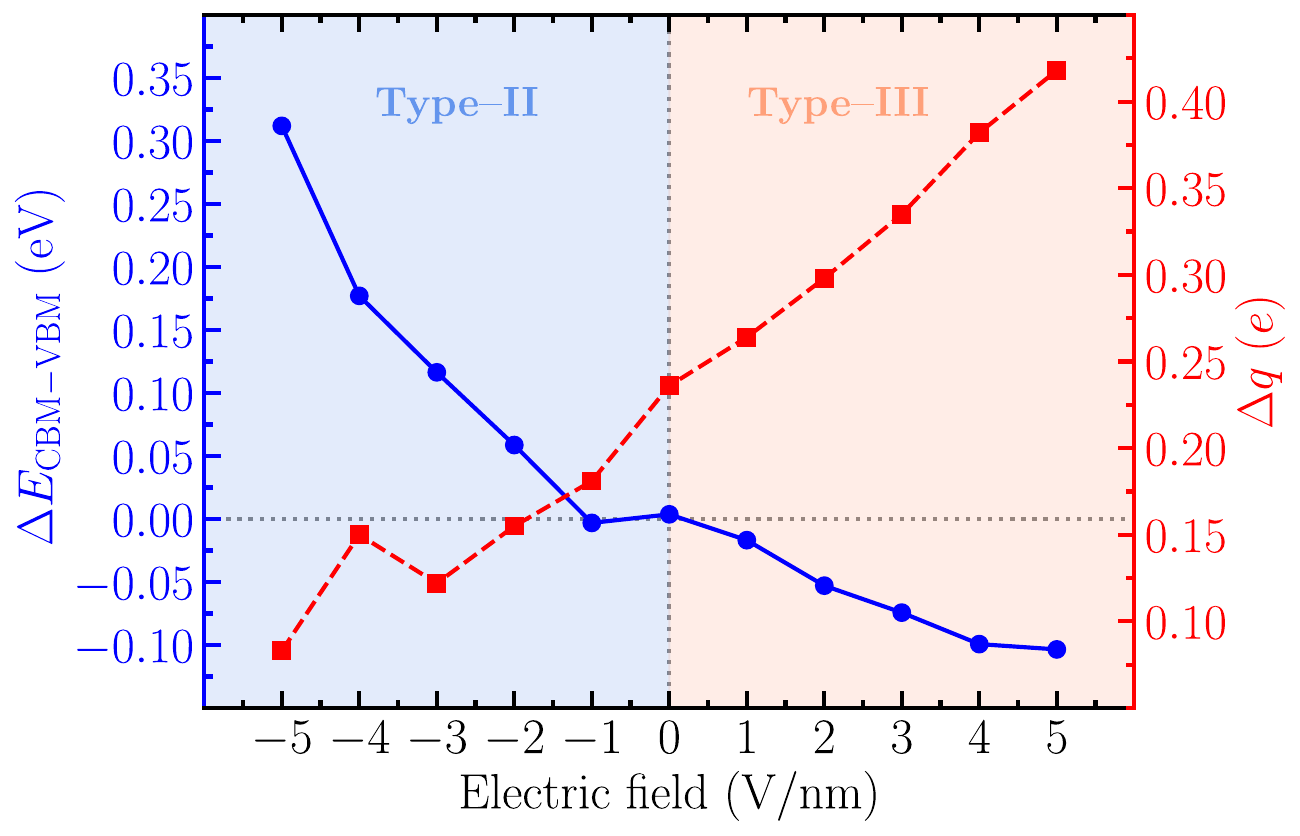}
\caption{(\textcolor{blue}{Color online}) Same as Fig. 6, but for $t-$MoTe$_2$/HfS$_2$.}
\label{fig:FIG9}
\end{figure}

In addition, the charge transfer $\Delta q$ increases monotonically with the external electric field. This trend reflects the progressive enhancement of interfacial charge redistribution induced by the applied field and is fully consistent with the increasing asymmetry observed in both the electronic band structures and the Rashba spin textures. For $E_{\mathrm{ext}}>0$, the larger charge transfer is directly associated with the strengthening of the type--III band alignment, which favors band-to-band tunneling and enhances the interlayer coupling. This behavior arises because the external electric field is aligned with the intrinsic electric field generated by interfacial charge transfer, thereby amplifying the charge separation across the interface and strengthening the electrostatic interaction between the two layers. Consequently, the increased interfacial charge density modifies the electronic transport properties of the heterostructure~\cite{Gunkel2020,Liang2025} and further enhances the overlap between the MoTe$_2$-derived VBM and the HfS$_2$-derived CBM, reinforcing the broken-gap character of the system.

The correlation between the increase in $\Delta q$, the enhanced CBM--VBM overlap, and the modulation of the spin texture demonstrates that the external electric field simultaneously affects the orbital, charge, and spin degrees of freedom of the vdWH. These results highlight the remarkable electrostatic tunability of the MoTe$_2$/HfS$_2$ heterostructure, enabling electrical control of both the transition between type--II and type--III band-alignment regimes and the Rashba spin splitting. Such characteristics are highly desirable for the development of low-power TFET and spintronic devices.

Figures~\ref{fig:FIG8}(a)--(e) present the electronic band structures of the twisted heterostructure $t$--MoTe$_2$/HfS$_2$ ($\theta = 21.79^\circ$) under different external electric fields. Similar to the untwisted system, the electric field strongly modifies the relative positions of the MoTe$_2$- and HfS$_2$-derived band edges. For negative electric fields, the HfS$_2$-derived CBM remains above the MoTe$_2$-derived VBM, corresponding to a type--II band alignment. In contrast, positive electric fields progressively reduce the energy difference $\Delta E_{\mathrm{CBM-VBM}}$, eventually driving the system into the type--III (\textit{broken-gap}) regime. Notably, this transition occurs in the vicinity of $E_{\mathrm{ext}} \approx 0$~V/nm, indicating that the twisted heterostructure lies intrinsically close to the type--II/type--III boundary.

The quantitative evolution of this behavior is summarized in Fig.~\ref{fig:FIG9}. The value of $\Delta E_{\mathrm{CBM-VBM}}$ decreases nearly monotonically with increasing electric field, changing from approximately $0.31$~eV at $E_{\mathrm{ext}}=-5$~V/nm to about $-0.10$~eV at $E_{\mathrm{ext}}=5$~V/nm. In particular, at zero electric field, $\Delta E_{\mathrm{CBM-VBM}}\approx 0.004$~eV, revealing that the twisted heterostructure is already positioned extremely close to the crossover between type--II and type--III band alignment. This result indicates that twist engineering alone can substantially modify the electronic structure, bringing the system to the verge of a broken-gap configuration.

In addition, the Bader charge transfer $\Delta q$ increases significantly with the applied electric field, reflecting an enhanced interfacial charge redistribution. For positive fields, the larger $\Delta q$ is accompanied by a stronger type--III character, which increases the overlap between the MoTe$_2$-derived VBM and the HfS$_2$-derived CBM. These findings suggest that the combination of twist-angle engineering and electrostatic gating provides an additional degree of freedom for simultaneously controlling the band alignment, charge transfer, and interlayer coupling in $t$--MoTe$_2$/HfS$_2$. Consequently, twisted MoTe$_2$/HfS$_2$ heterostructures emerge as attractive platforms for tunable tunneling electronic devices.

\subsection{Electronic properties under an out-of-plane and in-plane strain}\label{sec.strain}

Strain engineering has emerged as an effective strategy for tuning the electronic properties of two-dimensional (2D) materials and van der Waals heterostructures (vdWHs)~\cite{He2013,Peng2020}. Unlike conventional bulk materials, 2D systems can sustain relatively large deformations without structural failure, allowing substantial modifications of the electronic band structure, interlayer coupling, and charge transfer even under moderate strain~\cite{CastellanosGomez2013}. In particular, both out-of-plane and in-plane strain have been shown to efficiently control the band alignment and electronic properties of TMD-based vdWHs~\cite{Lei2019,PhysRevMaterials.6.084001,Mao2022,Iordanidou2023,GUO2023155644}. Motivated by these findings, we investigate the effects of out-of-plane and in-plane strain on the electronic properties of the MoTe$_2$/HfS$_2$ heterostructures, considering both the untwisted configuration and the twisted system with a twist angle of $\theta = 21.79^\circ$.
\begin{figure*}[!t]
\includegraphics[width=0.95\linewidth]{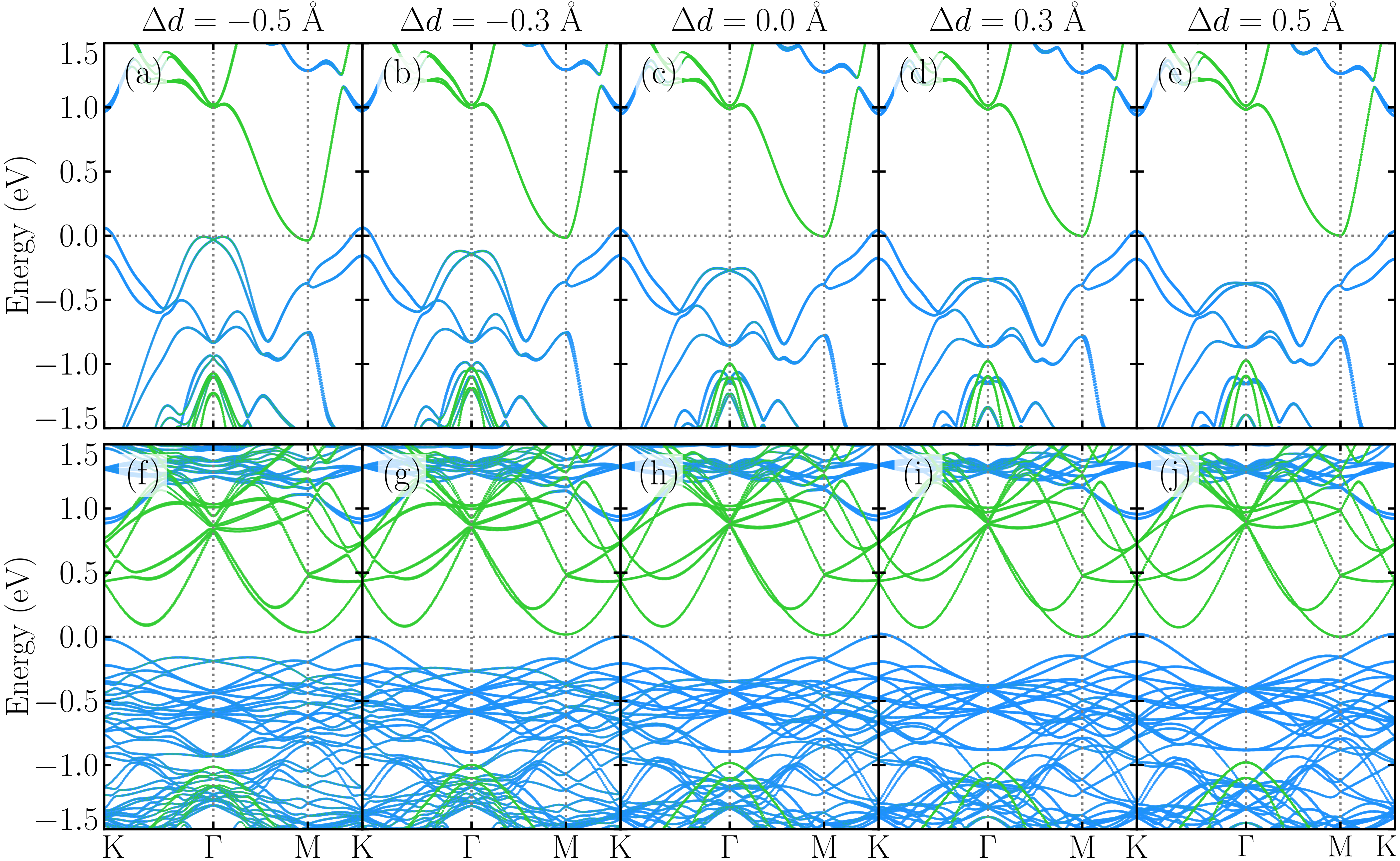}
\caption{(\textcolor{blue}{Color online}) Electronic band structures of the MoTe$_2$/HfS$_2$ vdWH under out-of-plane strain $\Delta d$, calculated using the PBE$+$SOC approach. Panels (a)--(e) correspond to the non-rotated heterostructure, while panels (f)--(j) show the twisted heterostructure. Negative and positive values of $\Delta d$ represent compressive and tensile out-of-plane strain, respectively. The blue and green curves indicate the contributions from the MoTe$_2$ and HfS$_2$ layers, respectively.}
\label{fig:FIG10}
\end{figure*}

To simulate out-of-plane strain in both heterostructures, the $z$ coordinates of the interfacial atoms were constrained, namely the upper Te atoms of the bottom layer and the lower S atoms of the top layer. The in-plane coordinates of these atoms, together with the positions of all remaining atoms, were allowed to fully relax during the structural optimization. The equilibrium configuration was used as the reference structure, and the interlayer distance was varied within the range $\Delta d = \pm0.5$~\AA. Negative and positive values of $\Delta d$ correspond to compressive and tensile out-of-plane strain, respectively.

Figure~\ref{fig:FIG10} presents the electronic band structures of the heterostructures under out-of-plane strain. Figures~\ref{fig:FIG10}(a)--(e) show the results for the MoTe$_2$/HfS$_2$ vdWH. For all investigated values of $\Delta d$, the heterostructure retains its type--III band alignment. Under tensile strain ($\Delta d > 0$), the MoTe$_2$-derived VBM at the K point shifts slightly toward lower energies, whereas the HfS$_2$-derived CBM at the M point moves upward relative to the Fermi level. As a result, the magnitude of the CBM--VBM overlap decreases with increasing interlayer separation. In addition, the Rashba-split bands around the $\Gamma$ point become progressively less pronounced as the tensile strain increases.

In contrast, compressive strain ($\Delta d < 0$) produces the opposite effect. The MoTe$_2$-derived VBM shifts toward higher energies, while the HfS$_2$-derived CBM moves downward, approaching and eventually crossing the Fermi level for the largest compressive strains considered. Simultaneously, the Rashba-split bands undergo a substantial energy shift toward the Fermi level, reaching it for the maximum compression investigated. Consequently, the CBM--VBM overlap increases under compressive strain, strengthening the type--III (\textit{broken-gap}) character of the heterostructure. This behavior is expected to enhance the BTBT process by widening the tunneling window and increasing the number of available tunneling channels.

Figures~\ref{fig:FIG10}(f)--(j) present the electronic band structures of the $t$--MoTe$_2$/HfS$_2$ heterostructure under out-of-plane strain. For tensile strain ($\Delta d > 0$), a behavior similar to that observed for the untwisted MoTe$_2$/HfS$_2$ vdWH is found. However,  the twisted heterostructure preserves its type--III (\textit{broken-gap}) band alignment throughout the entire strain range considered. At the same time, the magnitude of the CBM--VBM overlap gradually decreases as the interlayer separation increases.
\begin{figure*}[!t]
\includegraphics[width=2\columnwidth]{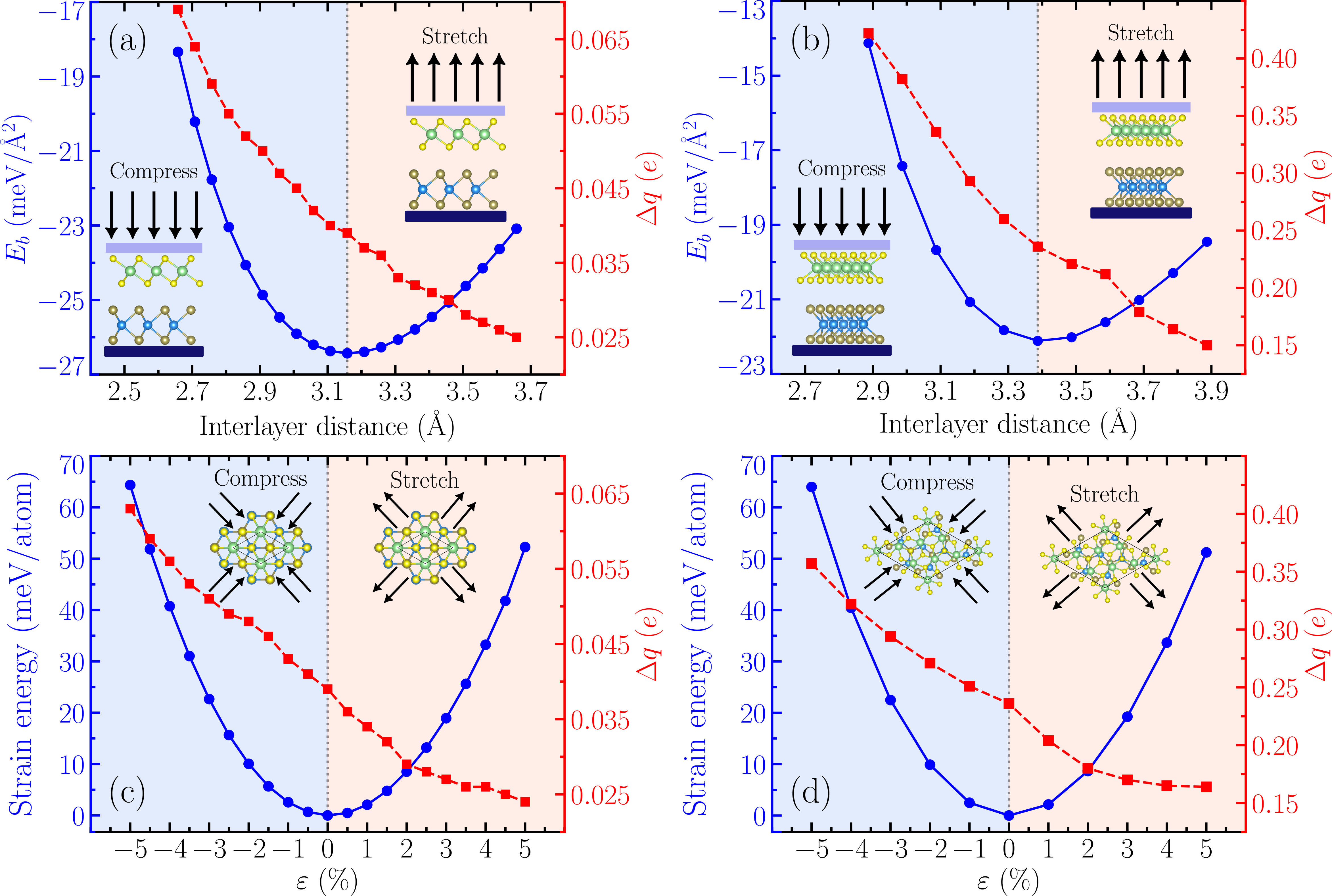}
\caption{(\textcolor{blue}{Color online}) Binding energy and Bader charge transfer of the MoTe$_2$/HfS$_2$ van der Waals heterostructures under out-of-plane and in-plane strain. Panels (a) and (b) show the dependence of the binding energy ($E_b$) and charge transfer ($\Delta q$) on the interlayer distance for the non-rotated and twisted ($\theta = 21.79^\circ$) heterostructures, respectively, under out-of-plane strain. Panels (c) and (d) display the strain energy and Bader charge transfer as a function of the biaxial in-plane strain $\varepsilon$ for the non-rotated and twisted heterostructures, respectively. The blue solid curves (left axis) represent the binding/strain energy, while the red dashed curves (right axis) correspond to the Bader charge transfer. The shaded blue and red regions indicate compressive and tensile strain regimes, respectively. The vertical dashed lines mark the equilibrium interlayer distance or the strain-free configuration. The schematic illustrations depict the compression and stretching processes applied perpendicular and parallel to the layers.}
\label{fig:FIG11}
\end{figure*}

A markedly different behavior is observed under compressive strain ($\Delta d < 0$). In this case, the MoTe$_2$-derived VBM at the K point shifts toward lower energies, whereas the HfS$_2$-derived CBM at the M point moves upward relative to the Fermi level. As a consequence, the overlap between the CBM and VBM progressively decreases, eventually leading to the opening of a finite band gap. In addition, the Rashba-split bands undergo a substantial energy shift toward the Fermi level as the compression increases. These results are qualitatively consistent with previous studies on twisted TMD-based heterostructures exhibiting type--III band alignment~\cite{PhysRevMaterials.6.084001,Iordanidou2023}. Overall, the results indicate a gradual transition from a type--III to a type--II band alignment, demonstrating that compressive out-of-plane strain provides an efficient mechanism for tuning the electronic structure of the twisted heterostructure.

Figures~\ref{fig:FIG11}(a)--(b) show the binding energy ($E_b$) and the Bader charge transfer ($\Delta q$) as a function of out-of-plane strain for the MoTe$_2$/HfS$_2$ vdWH and the $t$--MoTe$_2$/HfS$_2$ heterostructure, respectively. In both systems, $\Delta q$ exhibits a monotonic behavior, indicating charge transfer from the MoTe$_2$ layer to the HfS$_2$ layer. Under compressive strain ($\Delta d < 0$), the charge transfer increases owing to the reduced interlayer separation, which enhances the electronic interaction and orbital hybridization between the two layers. In contrast, tensile strain ($\Delta d > 0$) weakens the interlayer coupling and consequently reduces the charge transfer. Furthermore, the negative values of $E_b$ throughout the investigated strain range confirm the structural stability of both heterostructures.

We now discuss the effects of biaxial strain on the electronic properties of the MoTe$_2$/HfS$_2$ vdWH and the $t$--MoTe$_2$/HfS$_2$ heterostructure. The biaxial strain is defined as $\varepsilon = (a-a_0)/a_0 \times 100\%$, where $a$ and $a_0$ denote the lattice parameters of the strained and equilibrium structures, respectively. Positive (negative) values of $\varepsilon$ correspond to tensile (compressive) strain, and strain values ranging from $\varepsilon = -5\%$ to $+5\%$ were considered. For each strain value, the lattice parameters were fixed, while all internal atomic coordinates were fully relaxed until the convergence criteria were satisfied.
\begin{figure*}[!t]
\includegraphics[width=0.95\linewidth]{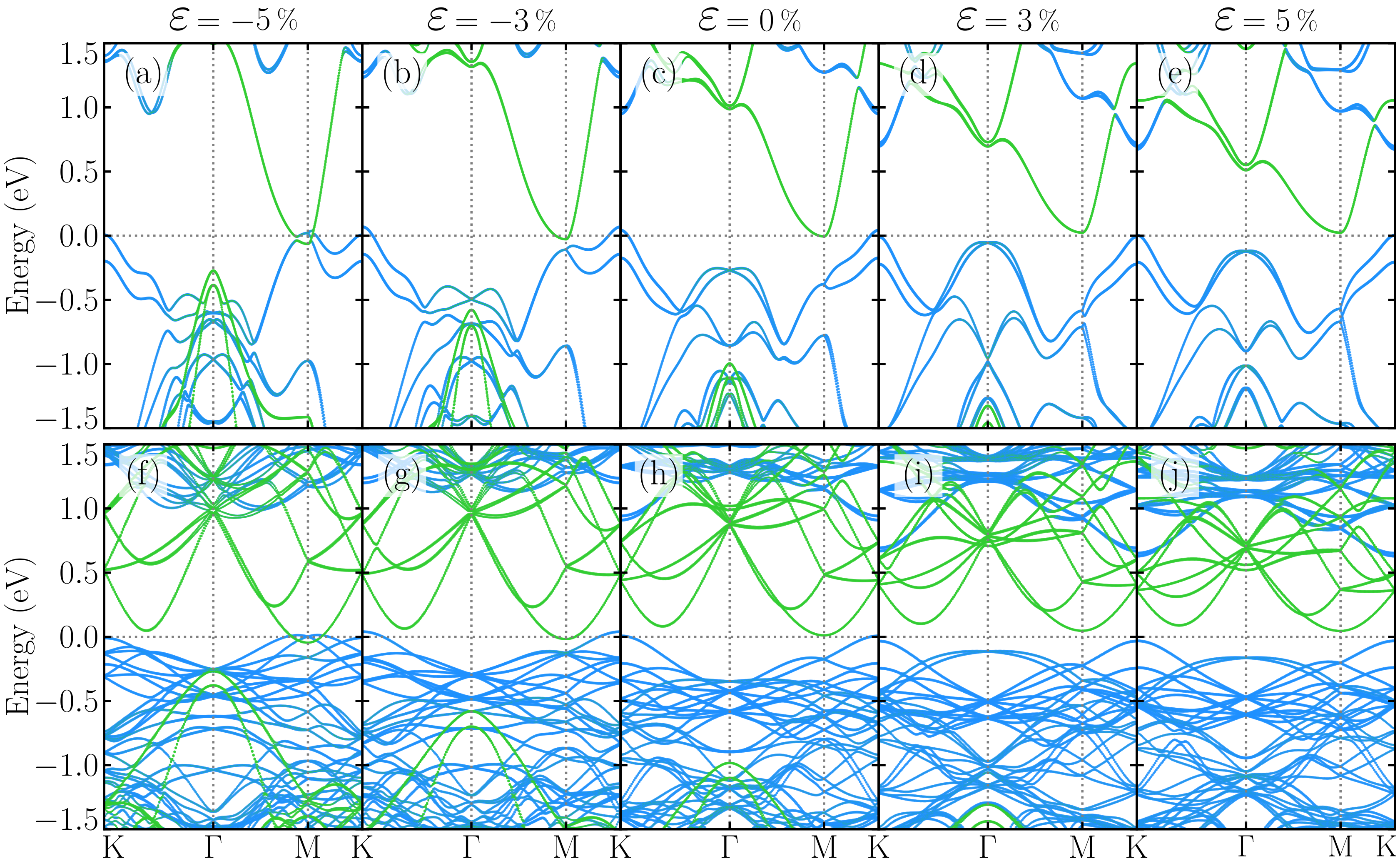}
\caption{(\textcolor{blue}{Color online}) Electronic band structures of the MoTe$_2$/HfS$_2$ vdWH under biaxial in-plane strain $\varepsilon$, calculated using the PBE$+$SOC approach. Panels (a)--(e) correspond to the non-rotated heterostructure, while panels (f)--(j) show the twisted heterostructure with twist angle $\theta = 21.79^\circ$. Negative and positive values of $\varepsilon$ represent compressive and tensile in-plane strain, respectively. The blue and green curves indicate the contributions from the MoTe$_2$ and HfS$_2$ layers, respectively.}
\label{fig:FIG12}
\end{figure*}

Figures~\ref{fig:FIG12}(a)--(e) present the electronic band structures of the MoTe$_2$/HfS$_2$ vdWH under different values of in-plane biaxial strain. As the tensile strain increases ($\varepsilon > 0$), the MoTe$_2$-derived VBM at the K point shifts slightly toward lower energies, while the HfS$_2$-derived CBM at the M point undergoes an energy shift that progressively reduces the overlap between the valence- and conduction-band edges. As a consequence, a finite band gap opens, driving a transition from the type--III (\textit{broken-gap}) alignment observed at $\varepsilon = 0\%$ to a type--II band alignment under tensile strain. This behavior indicates a gradual suppression of the BTBT probability as the tensile deformation increases.

In contrast, compressive biaxial strain ($\varepsilon < 0$) produces the opposite trend. The MoTe$_2$-derived VBM at the K point shifts toward higher energies, while the HfS$_2$-derived CBM at the M point is displaced in a manner that enhances the overlap between the band edges. For the largest compressive strains considered, the valence- and conduction-band states become strongly overlapped, reinforcing the type--III character of the heterostructure. Consequently, the tunneling window broadens and a larger number of electronic channels become available for transport, thereby favoring BTBT processes.

Figures~\ref{fig:FIG12}(f)--(j) display the electronic band structures of the $t$--MoTe$_2$/HfS$_2$ heterostructure under different in-plane biaxial strains. Overall, the twisted system exhibits a behavior qualitatively similar to that of the untwisted vdWH. Under tensile strain ($\varepsilon > 0$), the heterostructure undergoes a transition toward a type--II band alignment accompanied by the opening of a finite band gap. Conversely, under compressive strain ($\varepsilon < 0$), the type--III (\textit{broken-gap}) alignment is preserved and progressively strengthened due to the enhanced CBM--VBM overlap. These results demonstrate that biaxial strain provides an effective route for controlling the band alignment and tunneling characteristics of both the untwisted and twisted MoTe$_2$/HfS$_2$ heterostructures.

Consequently, for $\varepsilon > 0$, both heterostructures exhibit a type--II band alignment, in which electrons and holes tend to localize in different layers. Such spatial separation of charge carriers is particularly attractive for optoelectronic and photovoltaic applications, as it promotes the dissociation of photoexcited electron--hole pairs, suppresses carrier recombination, and can enhance the efficiency of energy-conversion processes. In contrast, for $\varepsilon < 0$, the strengthening of the type--III (\textit{broken-gap}) band alignment makes these heterostructures promising candidates for band-to-band tunneling devices, such as low-power tunnel field-effect transistors (TFETs).

The strain energy ($E_S$) was calculated according to $E_S=(E_{\mathrm{strain}}-E_{\mathrm{unstrain}})/N$, where $E_{\mathrm{strain}}$ and $E_{\mathrm{unstrain}}$ are the total energies of the strained and equilibrium structures, respectively, and $N$ is the total number of atoms in the heterostructure. Figures~\ref{fig:FIG11}(c)--(d) display the strain energy $E_S$ (blue curve, left axis) together with the Bader charge transfer $\Delta q$ (red curve, right axis), which quantifies the electron transfer from the MoTe$_2$ layer to the HfS$_2$ layer.

For both heterostructures, the strain energy exhibits an approximately quadratic dependence on $\varepsilon$, characteristic of elastic deformation. The charge transfer $\Delta q$ also displays a clear monotonic dependence on the applied strain. Under compressive strain ($\varepsilon < 0$), $\Delta q$ increases, indicating enhanced electron transfer from MoTe$_2$ to HfS$_2$. Conversely, as the strain shifts toward the tensile regime ($\varepsilon > 0$), the charge transfer gradually decreases. These trends are consistent with the evolution of the interlayer coupling and band alignment discussed above. Overall, the obtained results are in good qualitative agreement with previous studies on strained TMD--based van der Waals heterostructures~\cite{Lei2019,GUO2023155644,PhysRevMaterials.6.084001,Iordanidou2023,Iordanidou2024}.

\section{Conclusions}\label{sec.conclusions}

In this work, we performed a comprehensive first-principles investigation of the structural, electronic, charge-transfer, and spin-related properties of MoTe$_2$/HfS$_2$ van der Waals heterostructures, including the effects of stacking configuration, external electric fields, twist angle, and mechanical strain. Our calculations reveal that the most stable stacking configuration lies in close proximity to the type-II/type-III (\textit{broken-gap}) band-alignment boundary. Semilocal PBE$+$SOC calculations predict an intrinsic broken-gap alignment arising from the overlap between the MoTe$_2$-derived valence-band maximum and the HfS$_2$-derived conduction-band minimum, whereas HSE06$+$SOC calculations yield a small positive band gap, indicating that the heterostructure remains very close to the broken-gap regime. The spontaneous interlayer charge transfer generates an intrinsic electric field that breaks inversion symmetry and induces a pronounced Rashba spin splitting around the $\Gamma$ point, accompanied by a characteristic helical spin texture.

We further demonstrate that the electronic and spin properties of the heterostructure can be efficiently tailored by external perturbations. Vertical electric fields modulate the charge transfer, Rashba splitting, and band alignment, driving reversible transitions between type--II and type--III regimes. Twist engineering places the heterostructure close to the boundary between these two regimes, making its electronic structure highly sensitive to external stimuli. Mechanical strain provides an additional degree of control: out-of-plane and biaxial compressive strain enhance the broken-gap character and favor band-to-band tunneling, whereas tensile strain promotes a type--II alignment that is attractive for optoelectronic and photovoltaic applications.

Overall, our results establish MoTe$_2$/HfS$_2$ heterostructures as highly tunable multifunctional platforms in which charge transfer, band alignment, and Rashba spin splitting can be controlled in a coordinated manner through electrostatic gating, twist angle, and strain engineering. The coexistence of electrically tunable Rashba spin textures and controllable broken-gap electronic structures highlights the potential of these systems for next-generation low-power TFETs, spintronic devices, and multifunctional nanoelectronic technologies.

\section*{Acknowledgments} 
This work was partially supported by the Brazilian agencies CAPES, CNPq, and FAPDF. D.R.C. acknowledges support from CNPq (grants No. $312539/2025$--$8$,  No. $437067/2018$--$1$,  No. $423423/2021$--$5$,  No. $408144/2022$--$0$), the Research Foundation—Flanders (FWO), and the Fundação Cearense de Apoio ao Desenvolvimento Científico e Tecnológico (FUNCAP). A.C.D. acknowledges support from FAPDF (00193-00001817/2023-43, 00193-00002073/2023-84), CNPq (305174/2023-1, 444069/2024-0, 444431/2024-1), CENAPAD-SP/UNICAMP, the UNICAMP/FINEP-MCTI project, and the NACAD-UFRJ (Lobo Carneiro HPC). T.A.S.P. acknowledges support from the INCT on Materials Informatics (CNPq 371610/2023-0) and FAPEMAT (PRO-2025/00376). T.A.S.P. and A.C.D. also acknowledge PDPG-FAPDF-CAPES Centro-Oeste (00193-00000867/2024-94). This work used resources of the Centro Nacional de Processamento de Alto Desempenho at São Paulo (CENAPAD-SP).

\bibliography{zboxref/references}
\end{document}


\title{Rashba Effect and Type--III Broken--Gap Band Alignment in MoTe$_2$/HfS$_2$ van der Waals Heterostructures for Tunnel Field--Effect Transistors}

\author{Daniel S. Lopes\orcidlink{0000-0002-5075-1334}}
\email{daniel.lopes@fisica.ufc.br}
\affiliation{Department of Physics, University of Cear\'a, Fortaleza, CE, Brazil}
\affiliation{Institute of Physics, University of Mato Grosso, Cuiab\'a, MT, Brazil}

\author{Bruno P. e Silva\orcidlink{0000-0001-5830-8903}}
\email{bruno.poti@fisica.ufc.br}
\affiliation{Department of Physics, University of Cear\'a, Fortaleza, CE, Brazil}
\affiliation{Federal Institute of Education, Science and Technology of Cear\'a - \textit{campus} Camocim, Camocim, CE, Brazil}

\author{Diego R. da Costa\orcidlink{0000-0002-1335-9552}} 
\email{diego_rabelo@fisica.ufc.br}
\affiliation{Department of Physics, University of Cear\'a, Fortaleza, CE, Brazil}
\affiliation{Department of Physics, University of Antwerp, Antwerp, Belgium}

\author{Alexandre C. Dias\orcidlink{0000-0001-5934-8528}} 
\email{alexandre.dias@unb.br}
\affiliation{Institute of Physics and International Center of Physics, University of Brasília, Brasília, DF, Brazil}

\author{Rodrigo G. Amorim\orcidlink{0000-0001-9611-8772}}
\email{rgamorim@id.uff.br}
\affiliation{Department of Physics, Fluminense Federal University, Volta Redonda, RJ, Brazil}

\author{Teldo A. S. Pereira\orcidlink{0000-0001-7968-9194}} 
\email{teldo@fisica.ufmt.br}
\affiliation{Institute of Physics, University of Mato Grosso, Cuiab\'a, MT, Brazil}
\affiliation{National Institute of Science and Technology on Materials Informatics, Campinas, SP, Brazil}

\author{Jeanlex S. de Sousa\orcidlink{0000-0003-2836-0983}}
\email{jeanlex@fisica.ufc.br}
\affiliation{Department of Physics, University of Cear\'a, Fortaleza, CE, Brazil}

\maketitle

\tableofcontents

\section{Computational Technical Details}

\subsection{PAW Projectors}

\begin{table}[H]
\centering
\caption{\justifying{Basic information on the PAW projectors used in the simulations. The columns list, respectively: the chemical element, its atomic number, the corresponding POTCAR file, the number of valence electrons $Z_{\mathrm{val}}$, and the recommended minimum (ENMIN) and maximum (ENMAX) plane-wave cutoff energies.}}\label{tableS1}
\begin{threeparttable}
    \renewcommand{\arraystretch}{1.5} %
    \setlength{\tabcolsep}{4.5pt} %
\begin{tabular}{lccccc}
\toprule \toprule
Element  & Atomic number & POTCAR & $Z_{val}$ & ENMIN (eV) & ENMAX (eV)  \\ \midrule
\ce{Mo}  & 42   & Mo     & 6         & 168.438    & 224.584        \\
\ce{Te}  & 52   & Te     & 6         & 131.236    & 174.982        \\ 
\ce{Hf}  & 72   & Hf\_pv & 10        & 165.250    & 220.334        \\ 
\ce{S}   & 16   & S      & 6         & 194.016    & 258.689        \\ 
\specialrule{.1em}{.1em}{.2em} \Xhline{0.9pt}
\end{tabular}
\end{threeparttable}
\end{table}
\newpage
\subsection{Crystal Structures}

\vspace{0.5cm}
\begin{table}[htbp]
\centering
\caption{\justifying{Lattice parameters of bulk MoTe$_2$ and HfS$_2$ calculated with functionals (PBE, PBE+D2, PBE+D3, PBEsol, vdW-DF2 and rev-vdW-DF2) and their corresponding experimental values.}}
\label{tab:parametros}
    \begin{threeparttable}
    \renewcommand{\arraystretch}{1.5} %
    \setlength{\tabcolsep}{12pt} %
\begin{tabular}{lccc|ccc}
\toprule \toprule
 & \multicolumn{3}{c|}{MoTe$_2$} & \multicolumn{3}{c}{HfS$_2$} \\
\cmidrule(lr){2-4} \cmidrule(lr){5-7}
 & $a$ (\si{\angstrom}) & $c$ (\si{\angstrom}) & $V$ (\si{\angstrom}$^3$) & $a$ (\si{\angstrom}) & $c$ (\si{\angstrom}) & $V$ (\si{\angstrom}$^3$)\\
\midrule
Exp          & 3.520\tnote{$\S$}     & 13.910\tnote{$\S$}     & 149.260\tnote{$\S$} & 3.623\tnote{$\dagger$} & 5.841\tnote{$\dagger$} & 66.398\tnote{$\dagger$} \\
PBE          & 3.551     & 15.122     & 165.129 & 3.644 & 6.677 & 76.764 \\
PBE+D2       & 3.521     & 13.991     & 150.197 & 3.631 & 5.648 & 64.480 \\
PBE+D3       & 3.514     & 14.002     & 149.768 & 3.619 & 5.867 & 66.529 \\
PBEsol       & 3,490     & 13.952     & 147.198 & 3.587 & 5.913 & 65.891 \\
vdW$-$DF2      & 3.712     & 14.899     & 177.759 & 3.736 & 6.049 & 73.125 \\
rev$-$vdW$-$DF2  & 3.529     & 14.051     & 151.552 & 3.613 & 5.826 & 65.863 \\
\specialrule{.1em}{.1em}{.2em} \Xhline{0.9pt}
\end{tabular}
\label{table:lattice}
        \begin{tablenotes}
            \item[$\S$] Uesugi \textit{et al.}~\cite{Uesugi2017}.
             \item[$\dagger$] Rimmington  $\&$ Balchin~\cite{Rimmington1974}. 
        \end{tablenotes}
    \end{threeparttable}
\end{table}
\begin{figure}[H]
\includegraphics[width=0.75\linewidth]{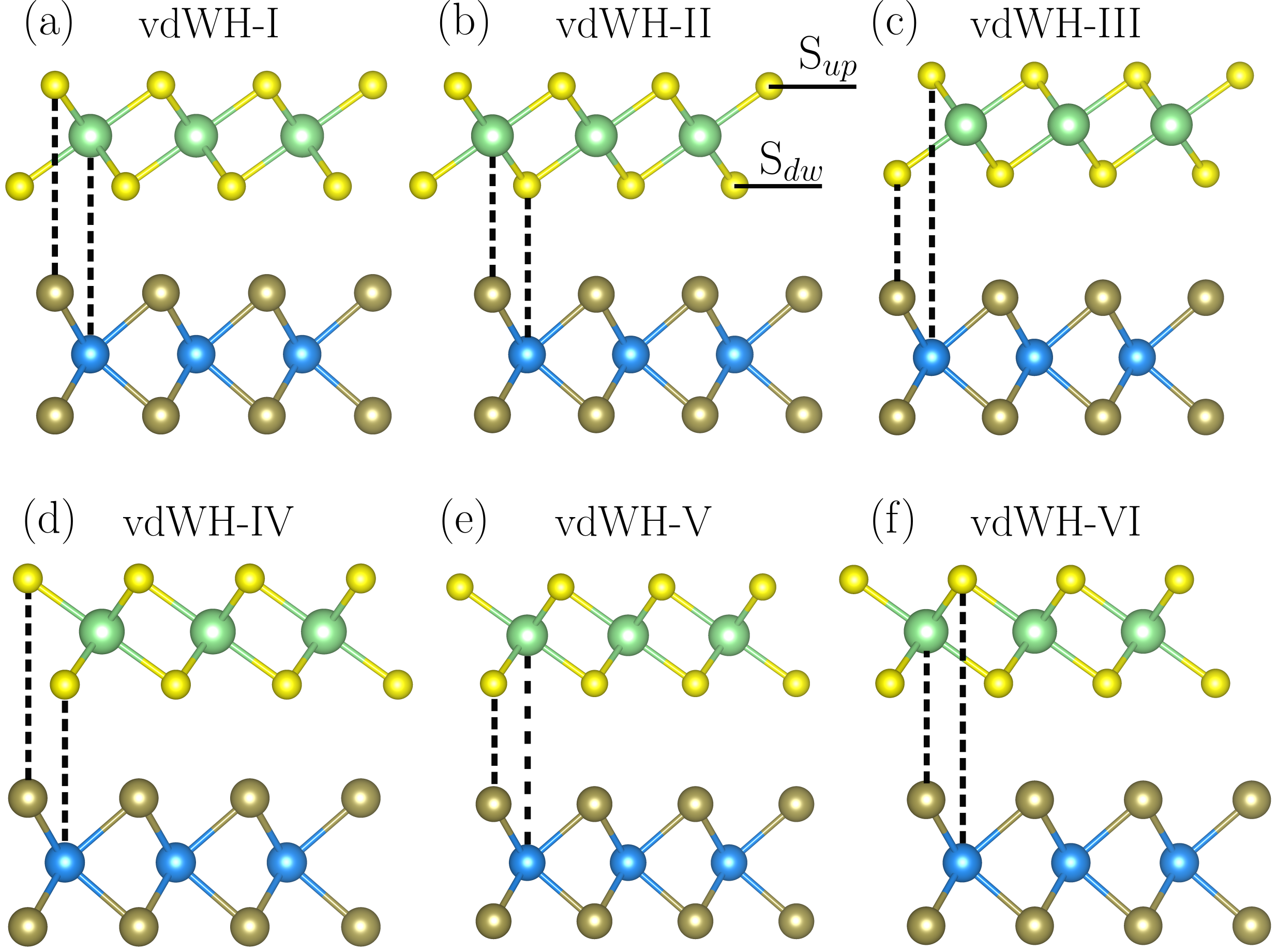}  
\begin{centering}  
\caption{(a)--(f) Atomic configurations of the MoTe$_2$/HfS$_2$ heterostructure for different high-symmetry stacking arrangements. The yellow, green, brown, and blue spheres denote S, Hf, Te, and Mo atoms, respectively.}
\end{centering}
\end{figure}
\newpage

The POSCAR file defines the crystal structure, including the lattice vectors, atomic positions, and atomic species corresponding to the stacking configurations investigated. This structural description serves as the basis for the analysis of the twisted MoTe2/HfS$_2$ heterostructure.
 
\begin{listing}[H]\scriptsize
\begin{Verbatim}
MoTe2/HfS2 (vdWH-I)                        
   1.00000000000000     
     3.5664300919000000    0.0000000000000000    0.0000000000000000
    -1.7832150459000000    3.0886190604000001    0.0000000000000000
     0.0000000000000000    0.0000000000000000   29.6311200240000048
  Mo              Te              Hf              S             
     1     2     1     2
Direct
     0.3333330000283397    0.6666669999547707    0.3985937124885197
     0.6666669999069086    0.3333329999157186    0.4589276407635055
     0.6666669999069086    0.3333329999157186    0.3380413488243406
     0.3333332697517051    0.6666665652621262    0.6136242582290379
     0.6666662697800447    0.3333335652168969    0.6634943584977933
     0.0000002696586137    0.9999995651778448    0.5637740997947134
\end{Verbatim}
    \caption{POSCAR relaxed of vdWH-I stacking.}
    \label{programa6}
\end{listing}

\begin{listing}[H]\scriptsize
\begin{Verbatim}
MoTe2/HfS2 (vdWH-II)                    
   1.00000000000000     
     3.5664300919000000    0.0000000000000000    0.0000000000000000
    -1.7832150459000000    3.0886190604000001    0.0000000000000000
     0.0000000000000000    0.0000000000000000   29.6311200240000048
  Mo              Te              Hf              S             
     1     2     1     2
Direct
     0.3333330000283397    0.6666669999547707    0.3962044552200581  
     0.6666669999069086    0.3333329999157186    0.4566671505449804  
     0.6666669999069086    0.3333329999157186    0.3356878526706737   
     0.6666665652714698    0.3333331305242524    0.6159586515569586  
     0.9999995652998095    0.0000001304790231    0.6658129085490998   
     0.3333335651783784    0.6666661304399710    0.5661244000561467 
\end{Verbatim}
    \caption{POSCAR relaxed of vdWH-II stacking.}
    \label{programa6}
\end{listing}

\begin{listing}[H]\scriptsize
\begin{Verbatim}
MoTe2/HfS2 (vdWH-III)                             
   1.00000000000000     
     3.5664300919000000    0.0000000000000000    0.0000000000000000
    -1.7832150459000000    3.0886190604000001    0.0000000000000000
     0.0000000000000000    0.0000000000000000   29.6311200240000048
  Mo              Te              Hf              S             
     1     2     1     2
Direct
     0.3333330000283397    0.6666669999547707    0.3880934460631522   
     0.6666669999069086    0.3333329999157186    0.4485967435240710   
     0.6666669999069086    0.3333329999157186    0.3275554273868835   
     0.0000000000000000    0.0000000000000000    0.6240483212737900   
     0.3333330000283397    0.6666669999547707    0.6739304264486066   
     0.6666669999069086    0.3333329999157186    0.5742310539014071   
\end{Verbatim}
    \caption{POSCAR relaxed of vdWH-III stacking.}
    \label{programa6}
\end{listing}

\begin{listing}[H]\scriptsize
\begin{Verbatim}
MoTe2/HfS2 (vdWH-IV)                   
   1.00000000000000     
     3.5664300919000000    0.0000000000000000    0.0000000000000000
    -1.7832150459000000    3.0886190604000001    0.0000000000000000
     0.0000000000000000    0.0000000000000000   29.6311200240000048
  Mo              Te              Hf              S             
     1     2     1     2
Direct
     0.3333330000283397    0.6666669999547707    0.3976429149820362  
     0.6666669999069086    0.3333329999157186    0.4580919721487646   
     0.6666669999069086    0.3333329999157186    0.3371280665306529  
     0.0000002697007275    0.0000005651792137    0.6145370652886797   
     0.6666662697800447    0.3333335652168969    0.6644003236079499   
     0.3333342696883221    0.6666675650918776    0.5646550760398412   
\end{Verbatim}
    \caption{POSCAR relaxed of vdWH-IV stacking.}
    \label{programa6}
\end{listing}

\begin{listing}[H]\scriptsize
\begin{Verbatim}
MoTe2/HfS2 (vdWH-V)                           
   1.00000000000000     
     3.5664300919000000    0.0000000000000000    0.0000000000000000
    -1.7832150459000000    3.0886190604000001    0.0000000000000000
     0.0000000000000000    0.0000000000000000   29.6311200240000048
  Mo              Te              Hf              S             
     1     2     1     2
Direct
     0.3333330000283397    0.6666669999547707    0.3886191457614032   
     0.6666669999069086    0.3333329999157186    0.4491038828015661   
     0.6666669999069086    0.3333329999157186    0.3280660881065245   
     0.3333335394524255    0.6666671304413399    0.6235065374399582   
     0.9999995395317427    0.0000001304790231    0.6734373578344020   
     0.6666675394400272    0.3333341303540038    0.5737224066540634   
\end{Verbatim}
    \caption{POSCAR relaxed of vdWH-V stacking.}
    \label{programa6}
\end{listing}

\begin{listing}[H]\scriptsize
\begin{Verbatim}
MoTe2/HfS2 (vdWH-VI)                           
   1.00000000000000     
     3.5664300919000000    0.0000000000000000    0.0000000000000000
    -1.7832150459000000    3.0886190604000001    0.0000000000000000
     0.0000000000000000    0.0000000000000000   29.6311200240000048
  Mo              Te              Hf              S             
     1     2     1     2
Direct
     0.3333330000283397    0.6666669999547707    0.3949377909865248   
     0.6666669999069086    0.3333329999157186    0.4554296925353469   
     0.6666669999069086    0.3333329999157186    0.3343728206108878   
     0.6666669999490225    0.3333339999170875    0.6172641752549950   
     0.3333330000283397    0.6666669999547707    0.6671105303122289  
     0.0000009999366171    0.0000009998297514    0.5673404088979339  

\end{Verbatim}
    \caption{POSCAR relaxed of vdWH-VI stacking.}
    \label{programa6}
\end{listing}
\begin{figure}[H]
\includegraphics[width=0.93\linewidth]{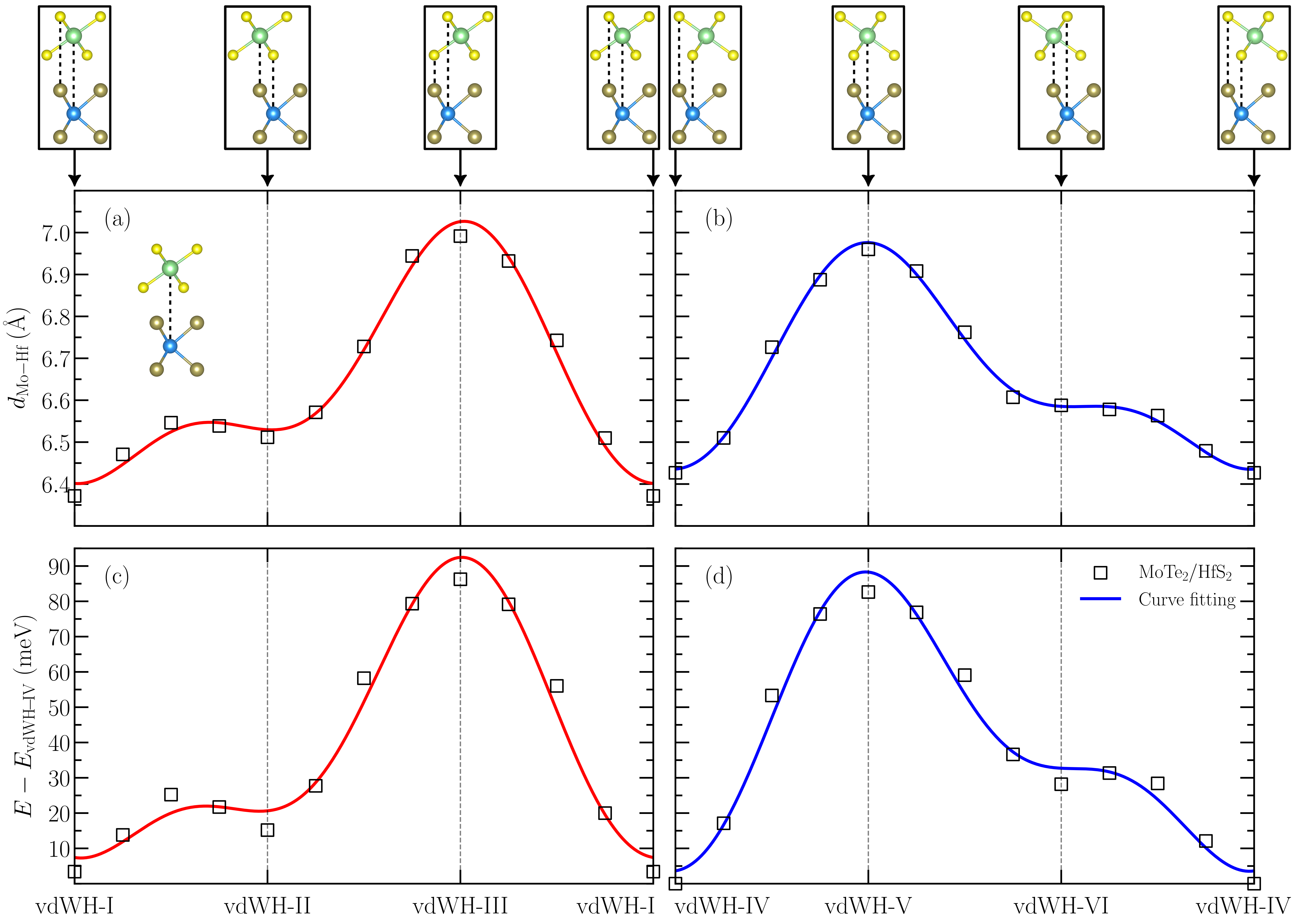}  
\begin{centering}  
\caption{(a)--(b) Interlayer distance $d_{\mathrm{Mo-Hf}}$ as a function of the high-symmetry stacking configurations for the MoTe$_2$/HfS$_2$ heterostructure. (c)--(d) Corresponding relative energy $E - E_{\mathrm{vdWH-IV}}$ for each stacking. Black squares represent the calculated data points, while solid lines denote the curve fitting. The dashed vertical lines indicate the different stacking configurations, illustrated schematically above each panel.}
\end{centering}
\end{figure}

\begin{table}[H]
\centering
\caption{\justifying{Calculated structural, energetic, and electronic properties of the MoTe$_2$/HfS$_2$ heterostructure for different high-symmetry stacking configurations (vdWH-I to vdWH-VI). The interlayer distance ($d_{\mathrm{int}}$), Mo--Hf vertical separation ($d_{\mathrm{Mo-Hf}}$), binding energy ($E_b$), and total energy are reported. The relative energy ($E - E_{\mathrm{vdWH-IV}}$) is referenced to the most stable configuration (vdWH--IV). The band alignment parameter ($\Delta E_{\mathrm{CBM-VBM}}$) is evaluated using the PBE, PBE+SOC, and HSE06 functionals. The charge transfer ($\Delta q$) is obtained from Bader charge analysis. The Rashba momentum offset ($k_R$) and Rashba energy ($E_R$) are also listed for each stacking configuration.}}
\begin{threeparttable}
    \renewcommand{\arraystretch}{1.28} %
    \setlength{\tabcolsep}{10pt} %
\begin{tabular}{lcccccc}
\toprule \toprule
 & vdWH-I & vdWH-II & vdWH-III & vdWH-IV & vdWH-V & vdWH-VI \\
\hline
$d_{\mathrm{int}}$ (\AA) & 3.107 & 3.243 & 3.723 & 3.158 & 3.693 & 3.316 \\
$d_{\mathrm{Mo-Hf}}$ (\AA) & 6.372 & 6.512 & 6.992 & 6.427 & 6.960 & 6.588 \\
$E_b$ (meV/\AA$^2$) & -26.121  & -25.052 & -18.649 & -26.430 & -18.929 & -23.872 \\
Total Energy (eV) & -42.219 & -42.207 & -42.137 & -42.222 & -42.140 & -42.194 \\
$E - E_{\mathrm{vdWH-IV}}$ (eV) & -3.412 & -15.185 & -85.721 & 0.000 & -82.635 & -28.183 \\
$\Delta E_{\mathrm{CBM-VBM}}$ (PBE) (eV) & 0.035 & 0.011 & 0.054 & 0.013 & 0.093 & -0.020 \\
$\Delta E_{\mathrm{CBM-VBM}}$ (PBE+SOC) (eV) & -0.027 & -0.047 & 0.023 & -0.048 & 0.017 & -0.082 \\
$\Delta E_{\mathrm{CBM-VBM}}$ (HSE06) (eV) & 0.271 & 0.228 & 0.324 & 0.234 & 0.392 & 0.122 \\
$\Delta q$ (e) & 0.042 & 0.035 & 0.028 & 0.039 & 0.029 & 0.034 \\
$k_R$ (\AA$^{-1}$) & 0.146 & 0.136 & 0.151 & 0.136 & 0.151 & 0.146 \\
$E_R$ (meV) & 21.216 & 16.492 & 9.088 & 17.256 & 9.665 & 15.874 \\
\specialrule{.1em}{.1em}{.2em} \Xhline{0.9pt}
\end{tabular}
        \label{table:lattice}
    \end{threeparttable}
\end{table}

\newpage
\begin{figure}[H]
\includegraphics[width=0.70\linewidth]{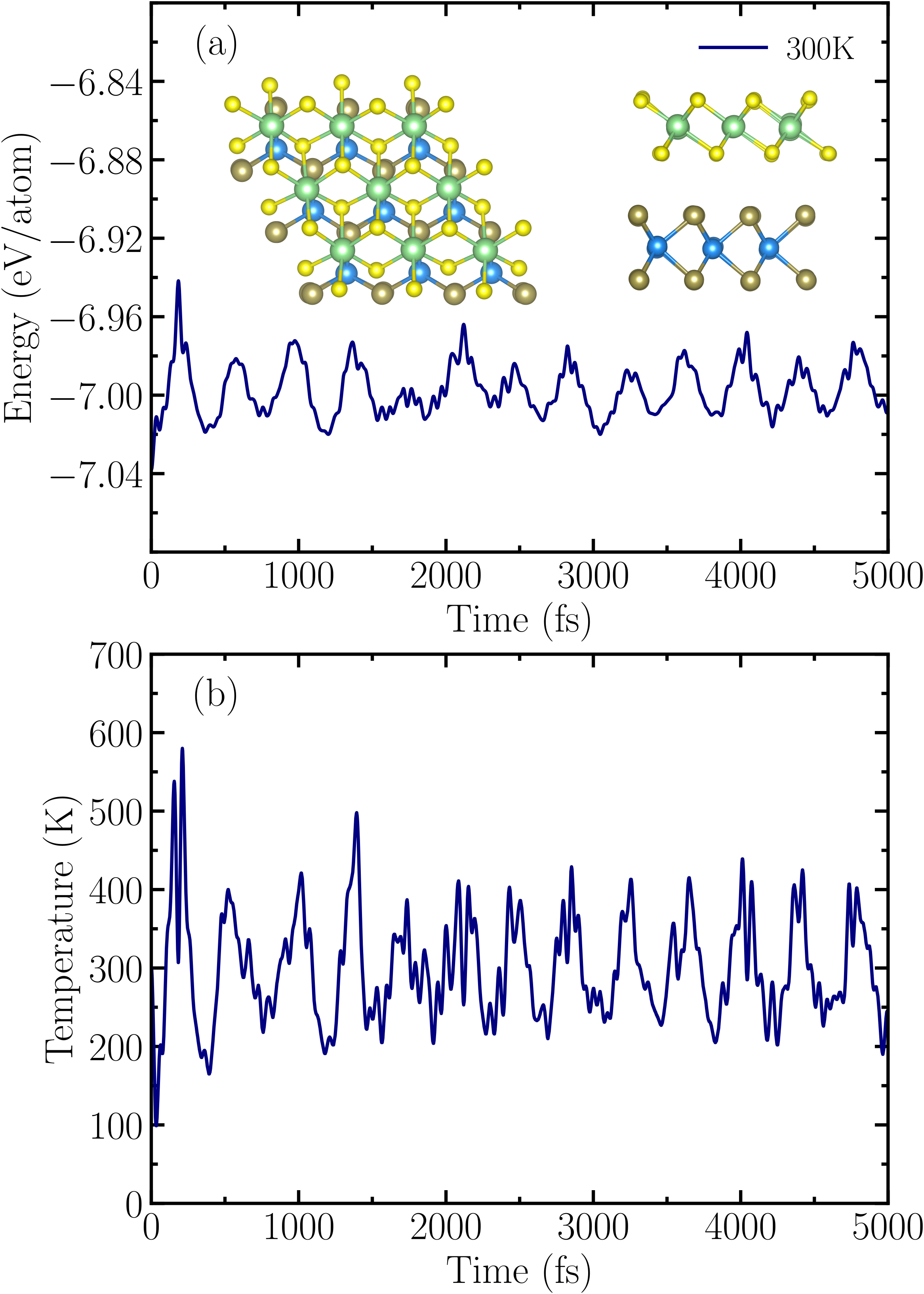}  
\begin{centering}  
\caption{
$Ab$ $initio$ molecular dynamics (AIMD) simulations of the MoTe$_2$/HfS$_2$ van der Waals heterostructure at 300 K. (a) Time evolution of the total energy per atom. (b) Temperature fluctuations throughout the simulation. The top and side views shown in the insets correspond to the final configuration after 5 ps, confirming the thermal stability of the heterostructure at room temperature.}
\end{centering}
\end{figure}

\newpage
\begin{figure}[H]
\includegraphics[width=0.95\linewidth]{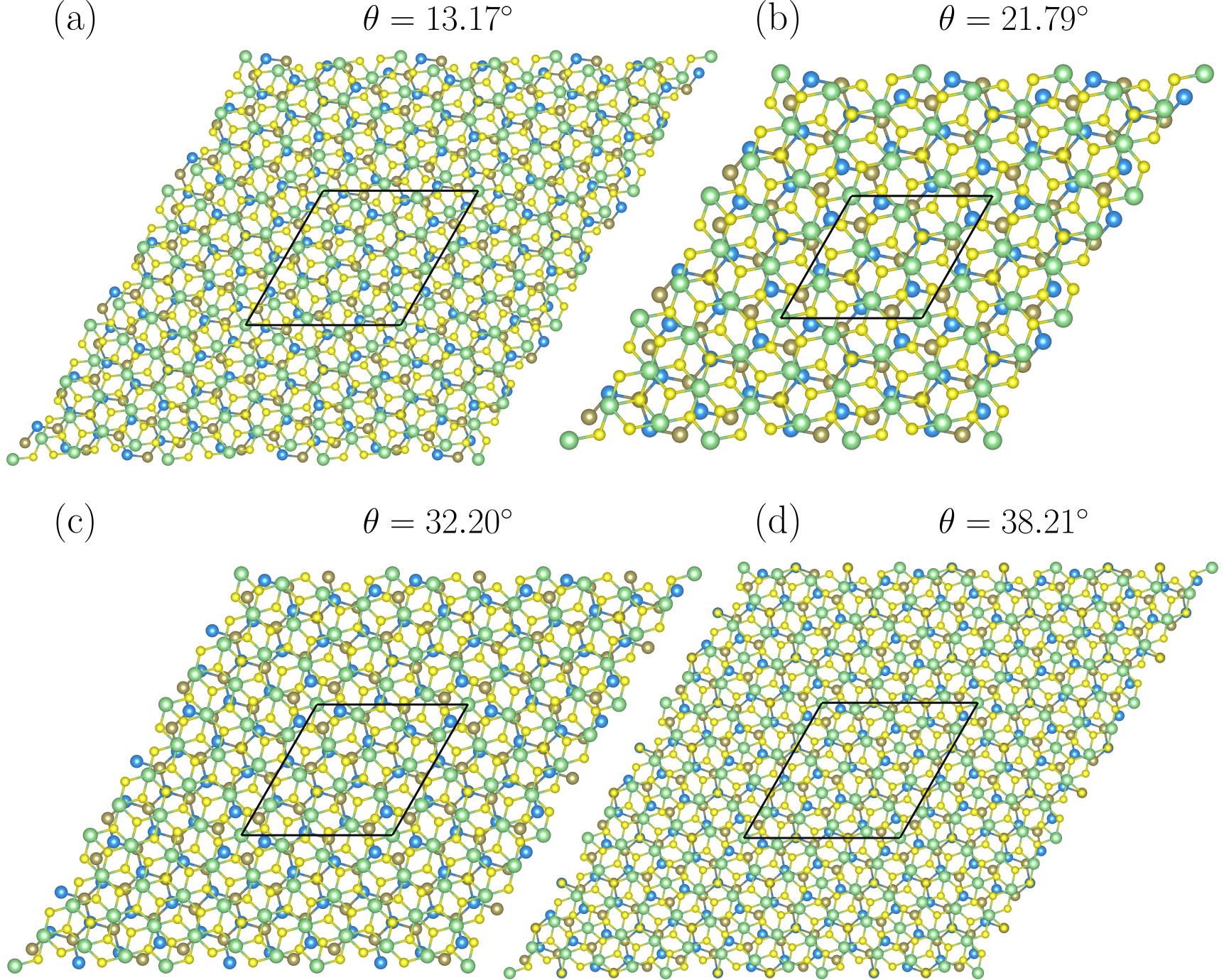}  
\begin{centering}  
\caption{Top views of the twisted MoTe$_2$/HfS$_2$ heterostructures ($t$-MoTe$_2$/HfS$_2$) for different commensurate twist angles: (a) $\theta = 13.17^\circ$, (b) $\theta = 21.79^\circ$, (c) $\theta = 32.20^\circ$, and (d) $\theta = 38.21^\circ$. The black rhombus in each panel highlights the corresponding moiré supercell. Yellow, green, brown, and blue spheres represent S, Hf, Te, and Mo atoms, respectively.}
\end{centering}
\end{figure}
\newpage
\section{Computational Details}\label{Sec.II}

All first-principles calculations were performed using the \textit{Vienna Ab Initio Simulation Package} (VASP)~\cite{Kresse1993,PhysRevB.49.14251,Kresse1996,KRESSE199615,PhysRevB.54.11169} within the projector-augmented wave (PAW) method~\cite{Blchl1994}. The atomic structures of the untwisted MoTe$_2$/HfS$_2$ van der Waals heterostructures were generated using the \textsc{pymatgen} library~\cite{ONG2013314}. Twisted heterostructures were constructed with the \textsc{Twister} package~\cite{Saismit2022}, which employs the Coincidence Lattice method~\cite{Koda2016, Uchida2014} approach to generate commensurate moiré supercells. In this method, the primitive lattice vectors of the two layers are chosen as

\begin{equation}
\begin{aligned}
\mathbf{a}_{1}^{(1)} &= \left(\frac{1}{2},\frac{\sqrt{3}}{2},0\right), &
\mathbf{a}_{2}^{(1)} &= \left(-\frac{1}{2},\frac{\sqrt{3}}{2},0\right),\\
\mathbf{a}_{1}^{(2)} &= \left(1,0,0\right), &
\mathbf{a}_{2}^{(2)} &= \left(-\frac{1}{2},\frac{\sqrt{3}}{2},0\right),
\end{aligned}
\end{equation}

\noindent where superscripts $(1)$ and $(2)$ denote the bottom and top layers, respectively. The coincidence supercell is defined by the lattice vectors

\begin{equation}
\mathbf{L}_{1}=m\mathbf{a}_{1}+n\mathbf{a}_{2},
\qquad
\mathbf{L}_{2}=-n\mathbf{a}_{1}+(m+n)\mathbf{a}_{2},
\end{equation}

\noindent with $(m,n)$ being integers specifying the commensurate supercell. The corresponding twist angle is given by

\begin{equation}
\cos (\theta_{m,n}) = \frac{n^{2} + 4nm + m^{2}}{2\,(n^{2} + nm + m^{2})},
\end{equation}

\noindent which generates all commensurate coincidence lattices for hexagonal systems. For a given pair $(m,n)$, the number of atoms per unit cell ($N_{\mathrm{atom}}$) is given by

\begin{equation}
        N_{\mathrm{atom}} = 4(n^{2}+nm+m^{2}).
\end{equation}

A vacuum region of 20~\AA\ was introduced along the out-of-plane direction to eliminate spurious interactions between periodic images.

For bulk MoTe$_2$ and HfS$_2$, several exchange--correlation functionals were employed to determine the structural parameters, including PBE~\cite{Perdew1996}, PBE$+$D2~\cite{Grimme2006,Kerber2008}, PBE$+$D3~\cite{Grimme2010}, PBEsol~\cite{Perdew2017}, vdW$-$DF2~\cite{Lee2010}, and rev$-$vdW$-$DF2~\cite{Hamada2014}. Based on these benchmark calculations, all subsequent simulations for the MoTe$_2$/HfS$_2$ vdWH and the twisted $t$--MoTe$_2$/HfS$_2$ heterostructure were performed using the PBE$+$D3 functional, which accounts for van der Waals interactions through semiempirical dispersion corrections.

A plane-wave kinetic-energy cutoff of 500~eV was adopted, corresponding to approximately $2\times$ENMAX, where ENMAX denotes the largest recommended cutoff energy among the constituent atomic species (see Table S1 in Section I). Structural relaxations and self-consistent field (SCF) calculations for the MoTe$_2$/HfS$_2$ vdWH ($t$--MoTe$_2$/HfS$_2$) were considered converged when the total-energy variation and residual forces on each atom were smaller than $10^{-7}$~eV ($10^{-5}$~eV) and $0.01$~eV/\AA\ ($0.01$~eV/\AA), respectively.

Brillouin-zone integrations were carried out using Monkhorst--Pack $\mathbf{k}$-point meshes corresponding to reciprocal-space densities of approximately $50~\text{\AA}^{-1}$ for structural relaxations and $60~\text{\AA}^{-1}$ for electronic-structure calculations. For the MoTe$_2$/HfS$_2$ vdWH, these densities correspond to $16\times16\times1$ and $19\times19\times1$ meshes, respectively. For the twisted heterostructure ($t$--MoTe$_2$/HfS$_2$), reciprocal-space densities of approximately $30~\text{\AA}^{-1}$ and $60~\text{\AA}^{-1}$ were employed for structural relaxations and electronic calculations, corresponding to $4\times4\times1$ and $8\times8\times1$ $\mathbf{k}$-point meshes, respectively. Spin--orbit coupling (SOC) was included in all electronic-structure calculations. To obtain a more accurate description of the electronic properties and to mitigate the well-known band-gap underestimation inherent to semilocal DFT, additional calculations were performed using the hybrid HSE06 functional~\cite{Krukau2006}.

For calculations involving external electric fields and planar-averaged electrostatic potentials, dipole corrections were applied along the out-of-plane direction. The thermal stability of the MoTe$_2$/HfS$_2$ heterostructure was further investigated by $ab$ $initio$ molecular dynamics (AIMD) simulations using a $3\times3\times1$ supercell and the same computational setup adopted for the twisted heterostructure. The simulations were performed within the canonical (NVT) ensemble at 300~K using the Nosé--Hoover thermostat~\cite{Nose10061984,Hoover1985}. The atomic trajectories were propagated for 5~ps with a time step of 1~fs.

Post-processing of the calculated data was carried out using the \textsc{VASPKIT}~\cite{WANG2021108033} and \textsc{PyProcar}~\cite{HERATH2020107080} packages, while atomic structures and charge-density visualizations were generated using the \textsc{VESTA} software~\cite{Momma:ko5060}.
\newpage

\section{Band structure}
\begin{figure}[H]
\includegraphics[width=0.95\linewidth]{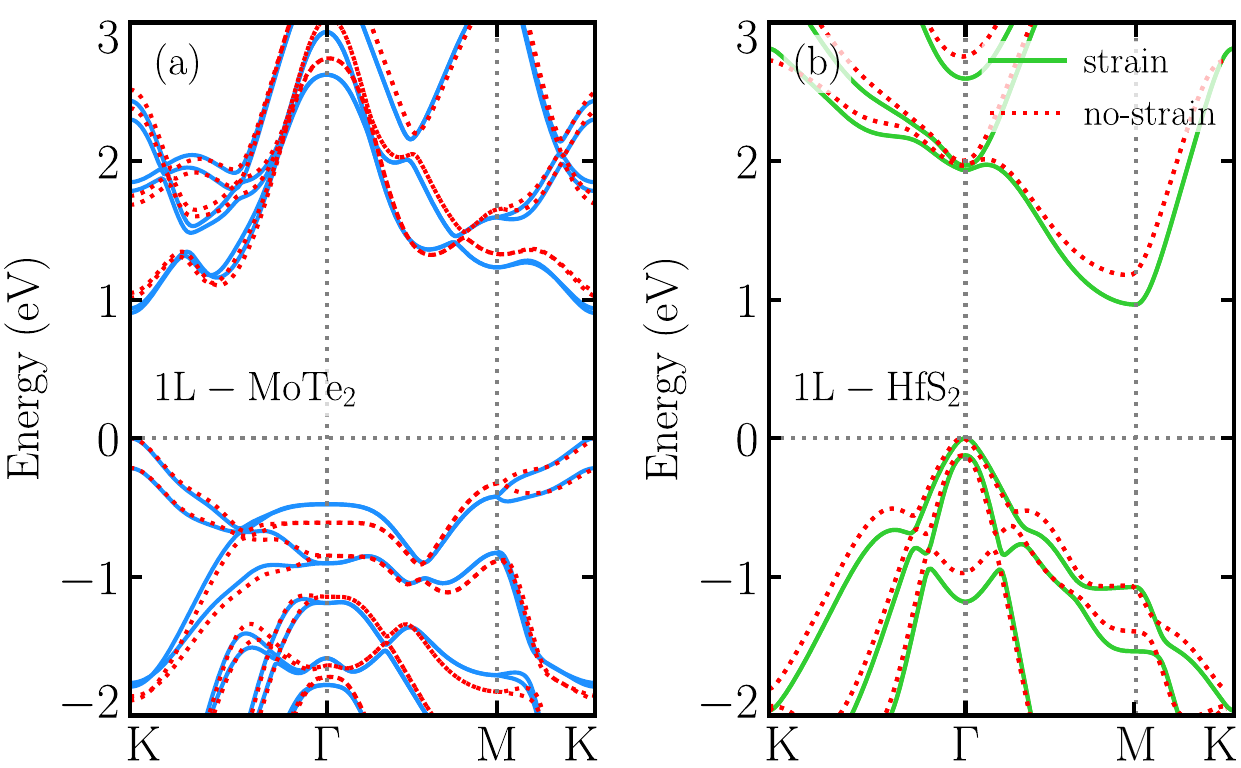}  
\begin{centering}  
\caption{Electronic band structures calculated using PBE$+$SOC for the isolated monolayers of (a) 1L--MoTe$_2$ and (b) 1L--HfS$_2$. The solid lines correspond to the monolayers under the small in-plane strain ($\sim$1.5\%) inherited from the formation of the MoTe$_2$/HfS$_2$ heterostructure, while the dotted lines represent the fully relaxed strain-free monolayers obtained using their optimized lattice parameters. The comparison shows that the applied strain induces only slight modifications in the electronic structure and band gap values.}
\end{centering}
\end{figure}
\newpage
\begin{figure}[H]
\includegraphics[width=1\linewidth]{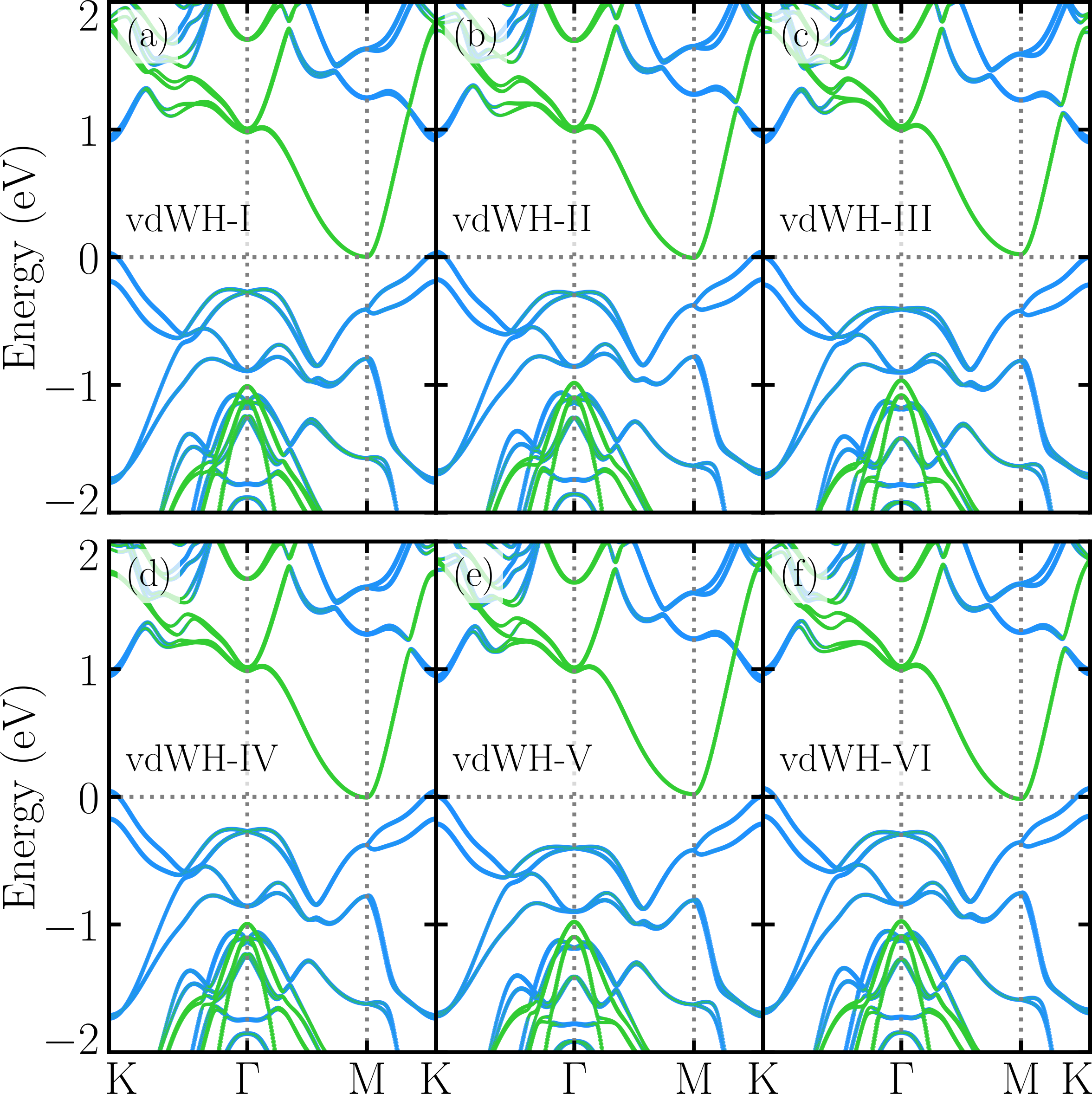}  
\begin{centering}  
\caption{Electronic band structures of the MoTe$_2$/HfS$_2$ heterostructure for different high-symmetry stacking configurations (vdWH-I to vdWH-VI), calculated using the PBE functional including spin–orbit coupling (PBE+SOC). The band structures are plotted along the high-symmetry path K-$\Gamma$–M–K. The blue and green curves represent the contributions from MoTe$_2$ and HfS$_2$ layers, respectively, highlighting the interlayer hybridization and band alignment for each stacking configuration.}
\end{centering}
\end{figure}
\newpage
\begin{figure}[H]
\includegraphics[width=1\linewidth]{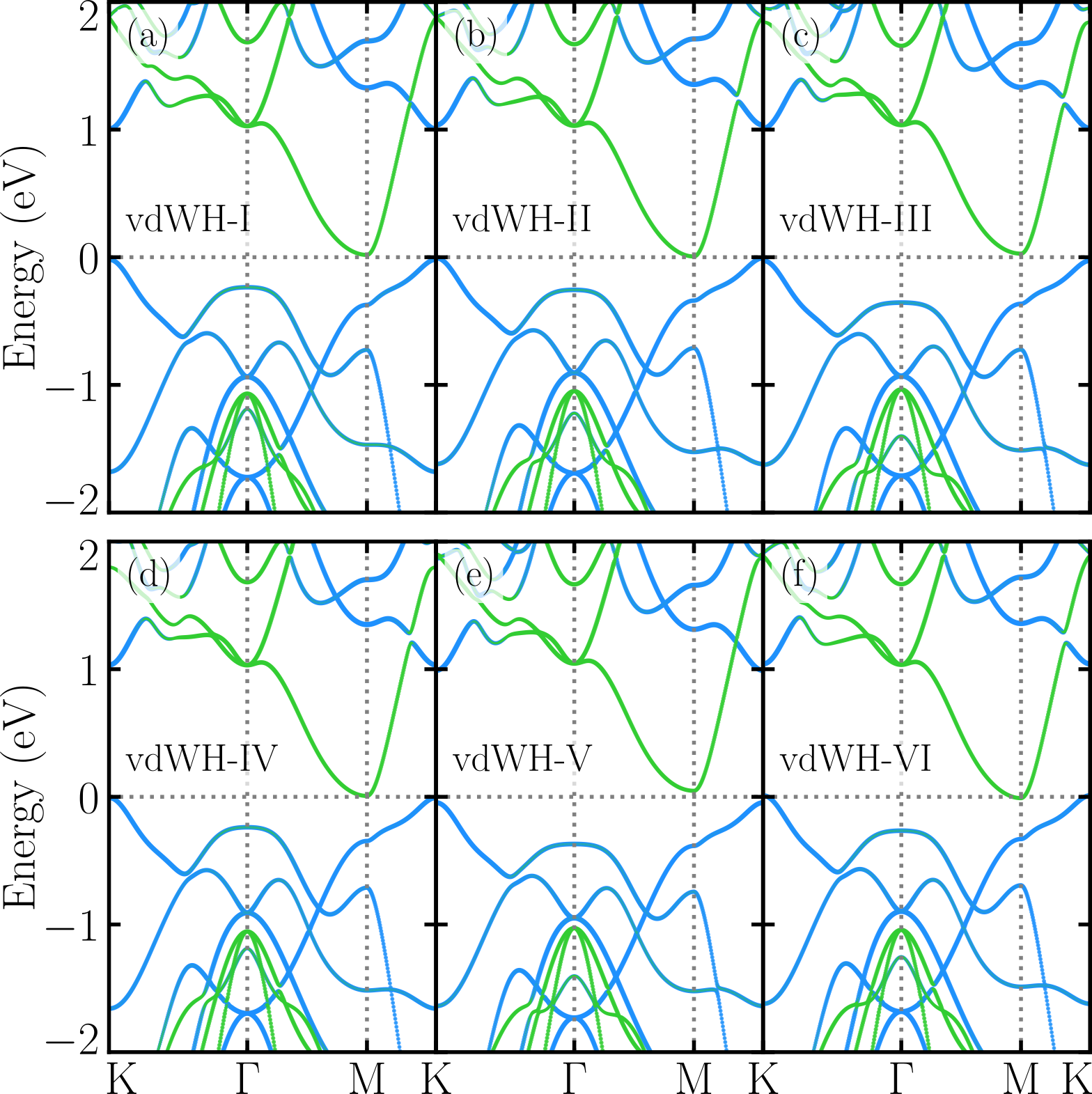}
\begin{centering}  
\caption{Electronic band structures of the MoTe$_2$/HfS$_2$ heterostructure for the six high-symmetry stacking configurations (vdWH-I to vdWH-VI), calculated within the PBE approximation without spin–orbit coupling (SOC). Compared to the PBE+SOC results, the overall band dispersion is preserved, while spin-related splittings are absent. This comparison highlights the role of spin–orbit coupling in modifying the band structure, particularly near the band edges, without significantly altering the global electronic features.}
\end{centering}
\end{figure}
\newpage
\begin{figure}[H]
\includegraphics[width=1\linewidth]{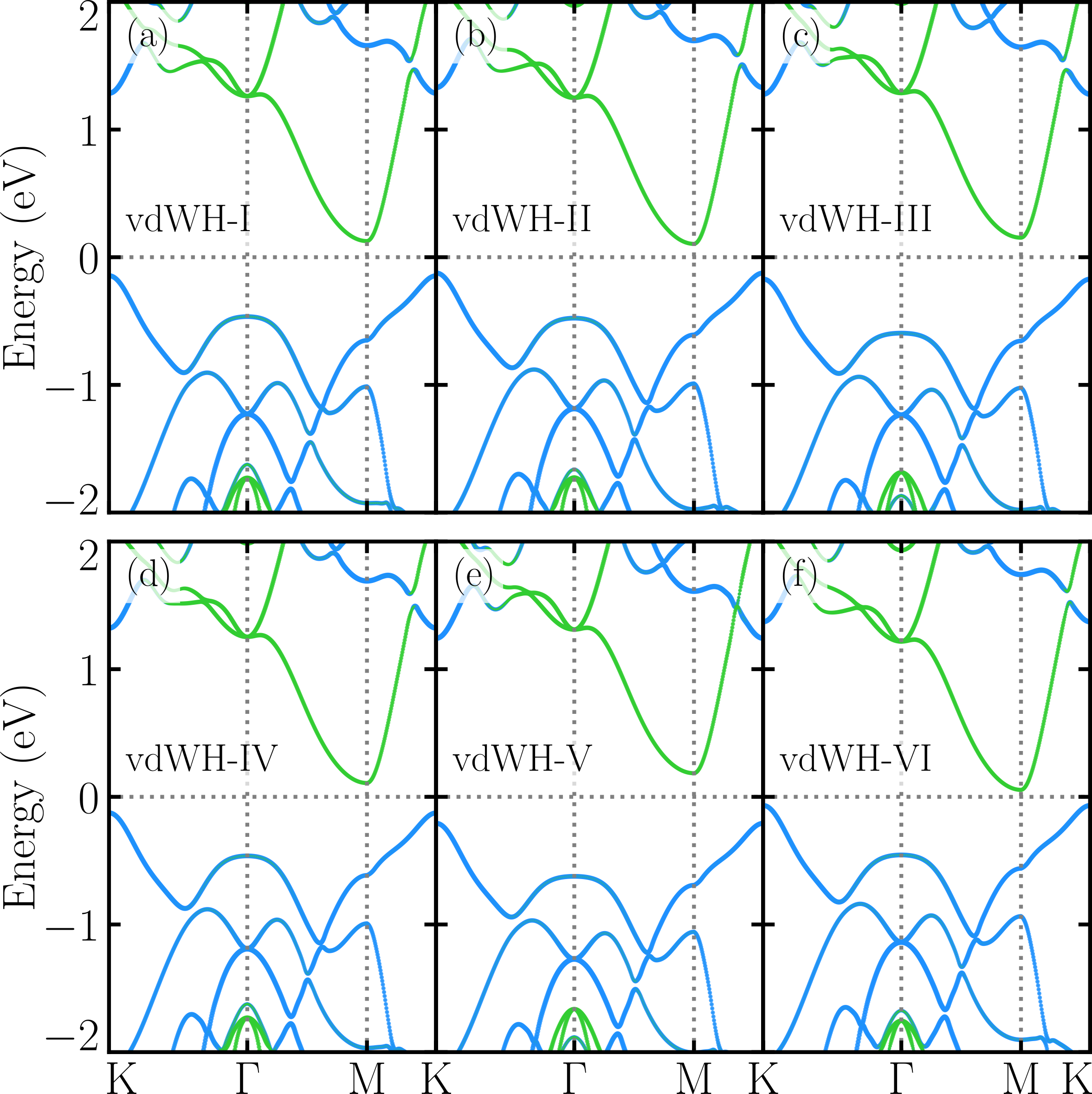}
\begin{centering}  
\caption{Electronic band structures of the MoTe$_2$/HfS$_2$ heterostructure for the six high-symmetry stacking configurations (vdWH-I to vdWH-VI), calculated within the HSE06 approximation without spin–orbit coupling (SOC).}
\end{centering}
\end{figure}

\begin{figure}[H]
\includegraphics[width=0.5\linewidth]{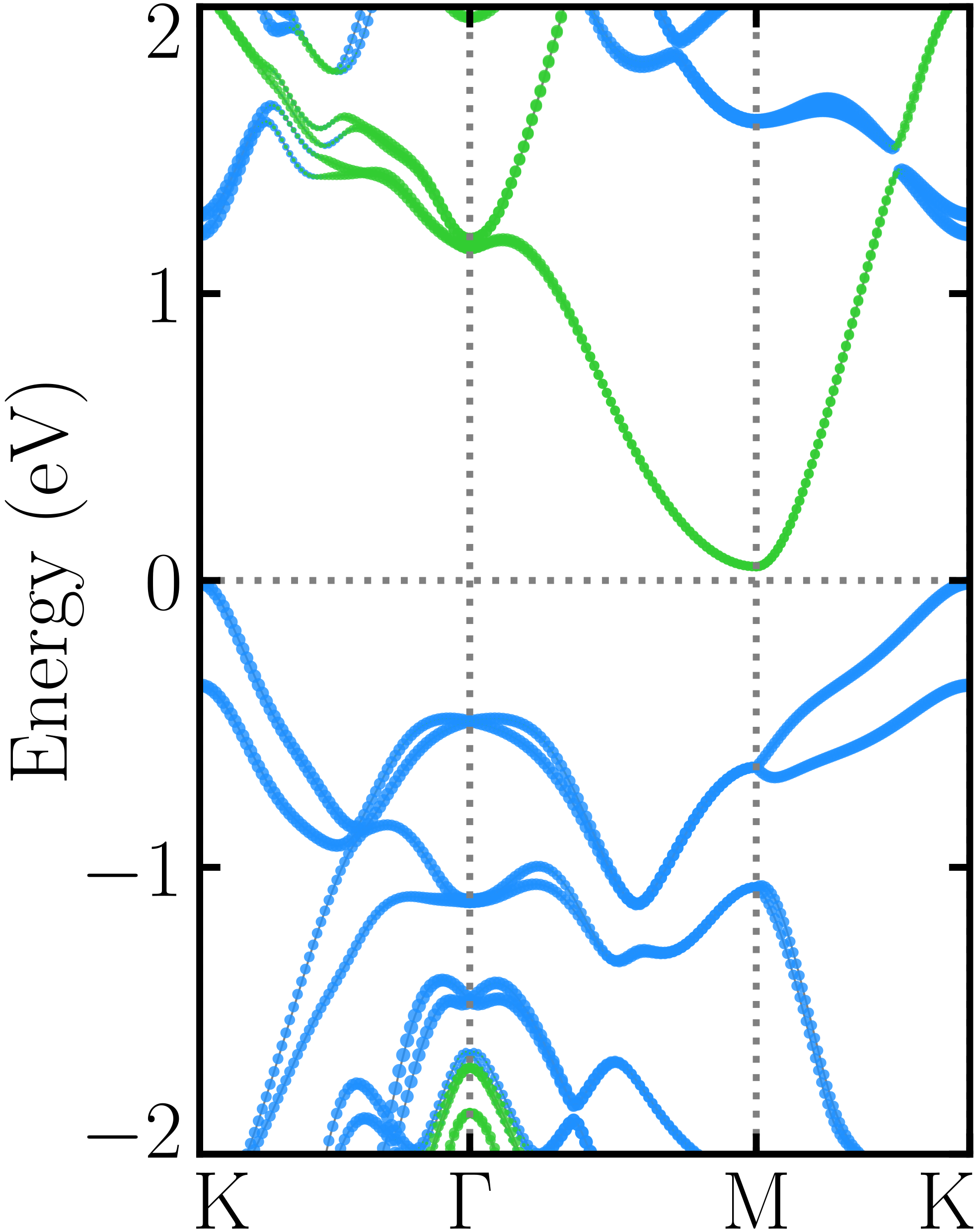}
\begin{centering}  
\caption{Electronic band structure of the energetically most stable MoTe$_2$/HfS$_2$ heterostructure (vdWH-IV), calculated using the HSE06 hybrid functional including spin--orbit coupling (HSE06+SOC). The orbital-projected bands highlight the contributions from the constituent layers, where the blue and green colors represent states predominantly derived from MoTe$_2$ and HfS$_2$, respectively.}
\end{centering}
\end{figure}
\newpage
\begin{figure}[H]
\includegraphics[width=1\linewidth]{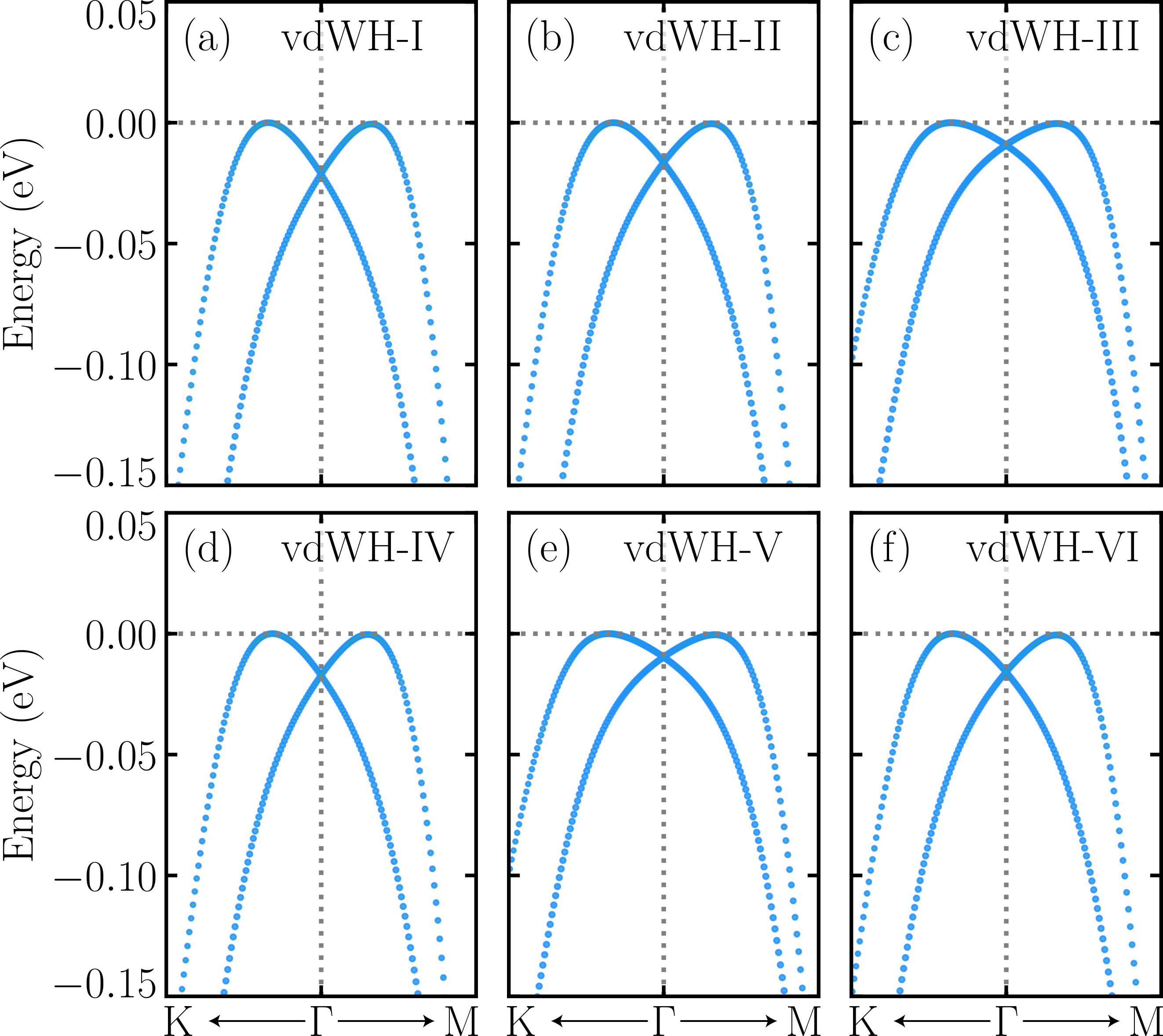}  
\begin{centering}  
\caption{Zoomed-in band structures around the $\Gamma$ point for the MoTe$_2$/HfS$_2$ heterostructure under different stacking configurations (vdWH-I to vdWH-VI), calculated using PBE including spin–orbit coupling (PBE+SOC). The valence bands exhibit a clear Rashba-type spin splitting, characterized by momentum-offset parabolic dispersions near $\Gamma$. The magnitude of the splitting and band curvature depend on the stacking configuration, reflecting variations in interlayer interaction and symmetry breaking.}
\end{centering}
\end{figure}
\newpage
\begin{figure}[H]
\includegraphics[width=0.8\linewidth]{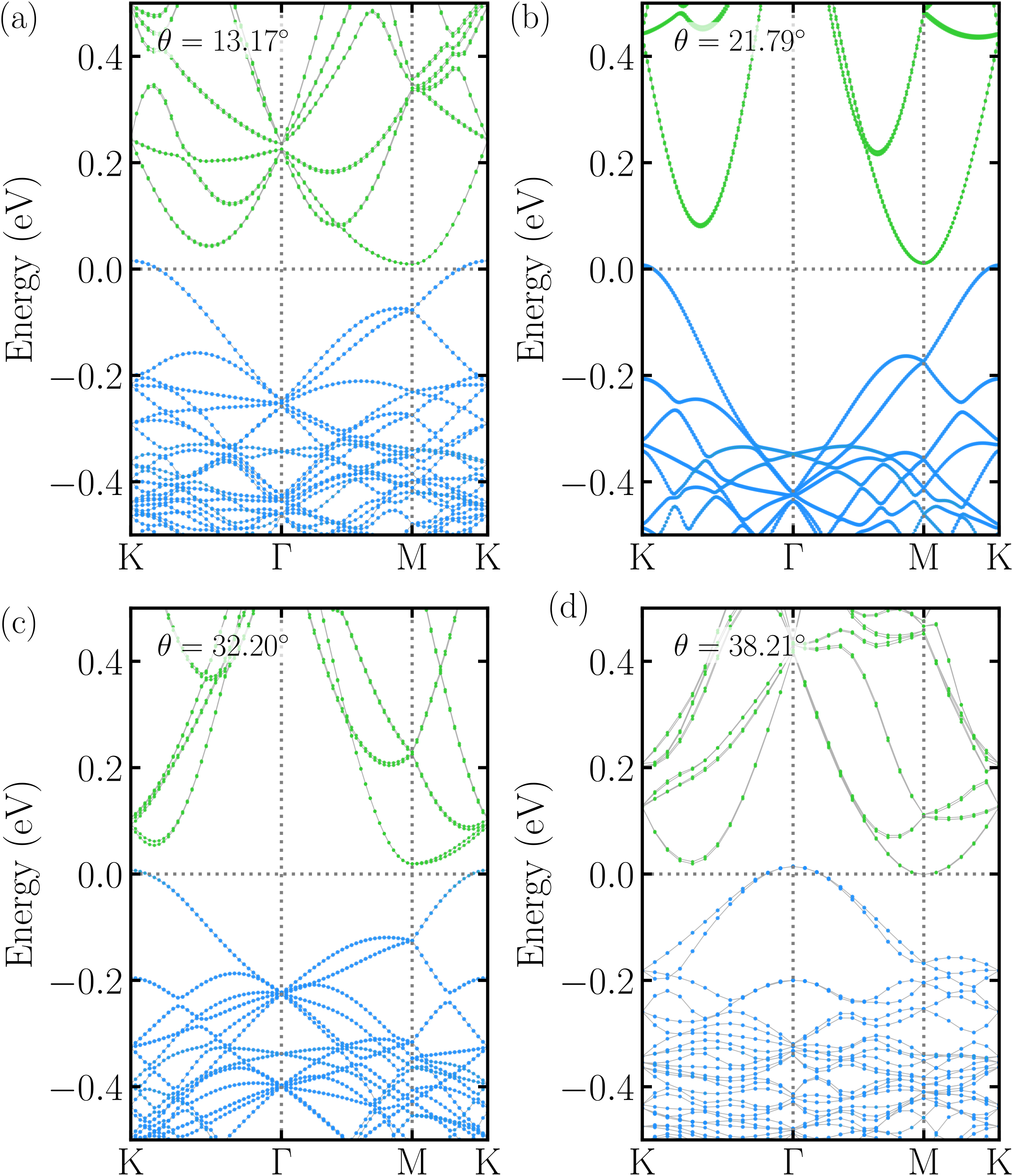}  
\begin{centering}  
\caption{PBE$+$SOC electronic band structures of twisted MoTe$_2$/HfS$_2$ heterostructures for twist angles of (a) $13.17^\circ$, (b) $21.79^\circ$, (c) $32.20^\circ$, and (d) $38.21^\circ$. Blue and green colors denote the projected contributions from the MoTe$_2$ and HfS$_2$ layers, respectively. The Fermi energy is set to zero. The band alignment remains governed by the MoTe$_2$ valence states and HfS$_2$ conduction states, while the relative CBM--VBM energy offset is strongly modulated by the twist angle.}
\end{centering}
\end{figure}
\newpage

\bibliography{zboxref/references}